\documentclass[12pt]{article}
\newlength{\abstractwidth}
\renewcommand{\thefootnote}{\fnsymbol{footnote}} 
\renewcommand{\thanks}[1]{\footnote{#1}} 
\newcommand{\starttext}{\setcounter{footnote}{0}
  \renewcommand{\thefootnote}{\arabic{footnote}}}

\makeatletter
\newcounter{fig}
\renewcommand\thefig{\arabic{fig}}

\def\fps@fig{tbp}
\def\ftype@fig{1}
\def\ext@fig{lof}
\def\fnum@fig{\figurename~\thefig}

\newenvironment{fig*}
               {\@dblfloat{fig}}
               {\end@dblfloat}

\makeatother

\usepackage{epsfig}

\newcommand{\be}{\begin{equation}}
\newcommand{\bea}{\begin{eqnarray}}
\newcommand{\eea}{\end{eqnarray}}
\newcommand{\ee}{\end{equation}}
\newcommand{\bd}{\begin{displaymath}}
\newcommand{\ed}{\end{displaymath}}

\newcommand{\<}{\langle}
\renewcommand{\>}{\rangle}
\def\ba{\begin{eqnarray}}
\def\ea{\end{eqnarray}}

\newcommand{\ud}{{\mathrm{d}}}

\newtheorem{theo}{theorem}

\def\eg{{\it e.g.~}}
\def\ie{{\it i.e.~}}

\def\Tr{{\rm Tr}}

\def\half{ {1\over 2}}

\def\la{\label}
\def\unit{1 \hskip-.3em \raise2pt\hbox{$ \scriptstyle |$ } }

\def\de{\delta}

\def\lm{\lambda}

\def\Si{\Sigma}

\def\G{\Gamma}

\def\O{\Omega}

\def\O{{\cal O}}

\def\og2{{\cal O}(g_2)}

\def\bop#1{\setbox0=\hbox{$#1M$}\mkern1.5mu
        \vbox{\hrule height0pt depth.04\ht0
        \hbox{\vrule width.04\ht0 height.9\ht0 \kern.9\ht0
        \vrule width.04\ht0}\hrule height.04\ht0}\mkern1.5mu}
\def\dg{\sp\dagger} 

\def\leftrighthookfill#1{$\mathsurround=0pt \mathord\hook#1
       \hrulefill\mathord\hook#1$}
\def\underhook#1{\vtop{\ialign{##\crcr                 
       $\hfil\displaystyle{#1}\hfil$\crcr
       \noalign{\kern-1pt\nointerlineskip\vskip2pt}
       \leftrighthookfill5\crcr}}}

\def\to{\rightarrow}

\makeatletter

\begin{document}

\begin{titlepage}

\leftline{\tt hep-th/0212118}

\vskip -.8cm

\rightline{\small{\tt CTP-MIT-3296}}

\begin{center}


{\LARGE\bf Predictions for PP-wave string amplitudes from perturbative SYM}
\vskip .3cm


 \vskip 1.cm

 {\large Umut G{\"u}rsoy}

\vskip 0.6cm

{\it Center for Theoretical Physics, \\
Laboratory for Nuclear Science and Department of Physics, \\
Massachusetts Institute of Technology, \\
Cambridge, Massachusetts 02139, USA} \\
E-mail: {\tt umut@mit.edu} \\

\end{center}

\vspace{1.7cm}

\begin{center}
{\bf Abstract}
\end{center}

The role of general two-impurity multi-trace operators in the BMN correspondence is
explored. Surprisingly, the anomalous dimensions of all two-impurity 
multi-trace BMN operators to order $g_2^2\lambda'$ are completely 
determined in terms of single-trace anomalous dimensions. 
This is due to suppression of connected field theory
diagrams in the BMN limit and this fact has important implications for
some string theory processes on the PP-wave background. 
We also make gauge theory predictions for the matrix elements of the
light-cone string field theory Hamiltonian in the two string-two
string and one string-three string sectors.

\noindent

\end{titlepage}

\newpage

\starttext

\section{Introduction}

PP-wave/SYM duality\cite{BMN} relates string states in IIB string
theory on the PP-wave background to a particular class of observables
in ${\cal N}=4$ SYM. These BMN operators 
are singled out in the Hilbert space of SYM by a
``modified'' 't Hooft limit,  
\begin{equation}\label{bmnlimit}
N\to\infty\textrm{, with }
\frac{J}{\sqrt{N}}\textrm{ and }g_{YM}~~\textrm{ fixed.}
\end{equation}
In detail, a single-trace BMN operator which involves two scalar
impurities is,
\be\la{BMNop}
O_n^J=\frac{1}{\sqrt{JN^{J+2}}}\sum_{l=0}^J\Tr(\phi Z^l\psi
Z^{J-l})e^{\frac{2\pi inl}{J}}.
\ee
In the {\emph{free}} string theory it is dual to a single-string state
with two-excitations,
$$\alpha^1_n\alpha_{-n}^2|p^+,0\>
$$where the light-cone momentum and R-charge $J$ are related by, 
$$\mu p^+\alpha '=\frac{J}{\sqrt{g_{YM}^2N}}\equiv \lm'.$$
Similarly an $i$-trace BMN operator is formed as,
\be\la{iBMN}
O_i^J=:O_n^{J_1}O^{J_2}\cdots O^{J_i}:\;,
\ee
from (\ref{BMNop}) and the BPS operators, 
\be\la{BPS}
O^J=\frac{1}{\sqrt{JN^J}}\Tr(Z^J),
\ee
In the {\emph{free}} string theory, this operator naturally maps into
an $i$-string state of the form,
$$\alpha^1_n\alpha_{-n}^2|p_1^+,0\>\otimes |p_2^+,0\>\otimes\cdots
|p_i^+,0\>.$$  

A striking aspect of PP-wave/SYM correspondence is that, in a regime
where both the effective GT coupling $\lm'$ and string coupling $g_s$
are small, 
one has {\emph{a duality between 
effectively weakly coupled gauge theory and perturbative string theory}}. 
This goes beyond the aforementioned duality between observables in SYM and string states on 
pp-wave background in the free string theory and provides an explicit map between gauge and string
interactions. However, a clear understanding of the correspondence at
the level of interactions still remains as an interesting challenge. One essential
reason which hinders a complete understanding is the fact that, while
states with different number of strings are orthogonal in SFT Hilbert
space at all orders in $g_s$, one gets a
non-trivial mixing between BMN operators of different number of traces when one
turns on the genus-counting parameter of SYM, $$g_2=\frac{J^2}{N},$$
which correspond to $g_s$ on the string side. Namely,
$$\<\bar{O}^J_iO^J_j\>_{g_2}\sim g_2^{|i-j|},\;\;\;\;\; i\ne j.$$ This is true
even in the free theory, \ie for $\lm'=0$. 

Recently, several important steps were taken in relation to this problem. A
natural route to take is to identify the {\emph{dynamical}} generators $P^-$
and $\Delta-J$ as,
$$\frac{2P^-}{\mu}=\Delta-J$$
also for non-zero values of $g_2$ and $g_s$ \cite{Gross2}. 
Since these operators act on completely different
Hilbert spaces, an unambiguous identification is achieved only by
equating the eigenvalues of $P^-$ and $\Delta-J$ in the corresponding
sectors of the Hilbert spaces. In case of one-string states this
problem was considered in a number of papers. On the gauge theory (GT) 
side, 
$\O(g_2^2)$ eigenvalue of BMN operators that correspond to
single-string states was obtained in
\cite{Gross2}\cite{German2}\cite{CFHM}. On the string theory side, 
one-string eigenvalue of $P^-$ at $\O(g_s^2)$ was first addressed
in \cite{PSVVV} where a computation that partially uses the language of 
String Bit Formalism (SBF) \cite{SBF} was performed 
and exact agreement with the GT result was reported. 
As noted in that paper however, an ultimate check of the
correspondence requires a purely string field theory (SFT) computation 
\cite{SFT}. Very recently, this calculation was carried out in
\cite{RSV} and also perfect agreement with GT eigenvalue 
was established.\footnote{Up to
  an ambiguity which arise from a particular truncation of the
  intermediate string-states.}  

Apart from the correspondence of eigenvalues, it is quite desirable to
have an identification of the {\emph{matrix elements}} of $P^-$ and
$\Delta-J$. This, of course requires, first to establish an isomorphism
between the complete bases that these elements are evaluated in. As
the BMN operators of different types mix with each other 
even in the free SYM, it becomes essential, first to 
give a characterization
of a particular ``string basis'' in GT in which BMN operators
$\tilde{O}_i$, at the free level form an orthonormal basis. 

One can fix the ambiguity in identification of the string
basis \eg by requiring the matrix elements of $\Delta-J$ at $\og2$ between
single and double-trace modified BMN operators match with
$\<\psi_1|P^-|\psi_2\>$ where $\psi_i$ denote an $i$-string
state. This question was addressed very recently by two seemingly
different methods, \cite{PSVVV} and \cite{Gomis} (whose compatibility
we demonstrate at $\O(g_2^2)$ in Appendix B.) 
After fixing the string basis at a certain order in $g_2$, evaluation of 
various matrix elements of $\Delta-J$ at this order
provides predictions for SFT computations. For example after fixing
the basis transformation at $\O(g_2^2)$ as explained above one should
be able to match the matrix elements
$\<\tilde{O}_i|\Delta-J|\tilde{O}_{i+2}\>$ with
$\<\psi_i|P^-|\psi_{i+2}\>$ whose leading order contributions are at
$\O(g_2^2)$.  
  
So far these problems were considered only in case of single and
double-trace BMN operators. In this note, we will take the first step
in addressing the role played by higher order multi-trace operators
in the perturbative PP-wave/SYM correspondence. We work out the case of
triple-trace BMN operators in detail. But we will be able to draw
some general conclusions which hold for {\emph{all}} two-impurity 
multi-trace BMN operators. 

Let us first address a subtlety in the 
determination $\Delta-J$ eigenvalue in GT which was discussed in
\cite{CFHM} and also relevant in the analysis of \cite{German2}. The mixing between BMN
operators of different number of traces indicates that $O_i$ does not have the required
transformation properties under dilatation for non-zero $g_2$. 
Therefore the first step in determination of the eigenvalue of
$\Delta-J$ is to construct an orthogonal set of BMN operators, 
$\tilde{O}_i$ which takes the effects of operator mixing into
account. Then one can obtain the eigenvalue $\Delta_i$
essentially by using the information contained in the free and
$\O(\lm')$ parts of the two-point function, $\<\bar{\tilde{O}}^J_i\tilde{O}^J_j\>$. 
In \cite{CFHM} it was pointed out that this diagonalization procedure is
equivalent to first-order non-degenerate perturbation theory where one
first obtains the eigen-operators of $\Delta-J$ at $\og2$, then
works out the $\O(g_2^2)$ eigenvalue as the second step in
perturbation theory. However to assure the validity of this procedure, one needs to establish the 
non-degeneracy at $\O(g_2^2)$. In \cite{German2}\cite{CFHM} it was found that the
non-degeneracy holds at $\og2$ thanks to very delicate
cancellations and the following result for the anomalous dimension was
obtained,
\be\la{contour}\Delta_{1}=
\frac{g_2^2}{4\pi^2}\left(\frac{1}{12}+\frac{35}{32\pi^2n^2}\right).
\ee
It was also pointed out that if perturbation theory
becomes degenerate at the next order in $g_2$, the result for the
eigenvalue would only be valid in the very special case of world-sheet
momenta $n=1$. In section 2, we perform the required analysis in
second-order perturbation theory by taking into account the mixing with 
the triple-trace operators and show that non-degenerate perturbation
theory becomes invalid at $\O(g_2^2)$. 
This phenomenon was first observed in \cite{German3}. This result casts some doubt on
the validity of eq. (\ref{contour}) for $n>1$ and necessiates the use 
of degenerate perturbation theory at $\O(g_2^2)$.
\footnote{In the first version of
  this paper we made the opposite conclusion due to an unfortunate
  error in section 2. We thank C. Kristjansen for pointing this out.}

However, as we briefly discuss at the end of section 2, one can show that the use
of degenerate in place of non-degenerate perturbation theory 
do not alter the previously obtained results drastically: 
For finite $J$, the degenerate subspace that contain the single-trace
operators consists of single-trace, triple-trace, 5-trace, \dots,
$J$=trace operators ($J$ is chosen as an odd number for
convenience). One can argue that there always exist an eigenstate in
this subspace which, in the limit, $J\to\infty$ continuously
transformed into $\tilde{O}_n^J$, and its eigenvalue tends to
eq. \ref{contour} in this limit. Futhermore there exist
eigenstates in the seperate degenerate subspaces of operators with odd
and even numbered traces whose eigenvalues in the BMN limit become the
anomalous dimensions associated with $\tilde{O_i^J}$.  
It is possible to demostrate this fact in a variety of effective
models. We would like to give details of this
interesting phenomenon along with new results associated with the use
of degenerate perturbation theory and the leave the general proof of the
aforementioned fact in a future work. Therefore our assertion can be 
taken as a conjecture in this paper. 

After establishing the validity of our method for the aforementioned 
particular eigenstates, one can ask for the
$\O(g_2^2)$ eigenvalues of higher trace BMN operators. We consider 
this problem in section 4. A rigorous investigation yields an
unexpected result: 
{\emph{Eigenvalues of all multi-trace BMN operators are solely
    determined by the eigenvalue $\Delta_1$ of single-trace
    operator as}}, 
$$\Delta_i =\left(\frac{J_1}{J}\right)^2\Delta_1.$$

This result is essentially due to the suppression of
connected GT correlators $\<\bar{\tilde{O}}_i\tilde{O}_j\>$ by a 
power of $J$ as $J\to\infty$ in the BMN limit. 
It is found that disconnected GT diagrams are less suppressed in this
limit and in fact only non-zero contributions to a generic correlator
of BMN operators arise from fully disconnected pieces. The connected
correlators will contribute to eigenvalues to higher order in $g_2$. 

Utilizing the correspondence of $\Delta-J$ with $P^-$ in the string
theory we show that this
fact translates into the absence of $\O(g_s^2)$ contact terms between 
states higher than single-string states. If the correspondence with
$P^-$ at the level of matrix elements holds, this also implies that
a particular class of tree-level string processes that would
contribute to the matrix elements on the PP-wave are
suppressed in the large $\mu$ limit. 
This conclusion is valid for processes in which the the external
string states that have two excitations along $i=1,2,3,4$ transverse
directions (that correspond to scalar impurities in
BMN operators).     

It is also interesting to investigate the the duality of $P^-$ and $\Delta-J$ at the
level of matrix elements. Using the method of \cite{Gomis} to fix the
basis transformation into ``string basis'' at $\O(g_2^2)$, we study
 the correlators of double and triple operators in this basis and
 obtain predictions for the matrix elements of $P^-$ at $\O(g_s^2)$, 
in double-double and single-triple sectors. These matrix elements are
given by remarkably simple expressions and solely determined by the
``non-contractible'' contribution to $\<\bar{O}_iO_j\>$ correlator
just as in the case of single-single matrix element
\cite{PSVVV}\cite{Gomis}. 
On the ST side, in the single-string sector the matrix element is
determined by the ''contact'' interaction between two single-string
states \cite{RSV}\cite{PSVVV}. Our study suggests a generalization of
this fact: a one-to-one map between non-contractible contributions to
GT correlation functions and contact interactions of the corresponding
states in SFT. These are explicit gauge theory results that are 
subject to check by a direct SFT calculation. 

We organize the paper as follows. In the next section we demonstrate  
the invalidity of non-degenerate perturbation theory in determining the
eigen-operators and eigenvalues of $\Delta-J$. Taking
into account the mixing with triple-trace operators we obtain the
mostly single-trace eigen-operator at $\O(g_2^2)$. 
We briefly outline our conjecture that use of degenerate and 
non-degenerate perturbation theory leads us to the same results
concerning the anomalous dimensions of particular eigen-states that
correspond to $\tilde{O}_n^J$ and $\tilde{O}_i^J$ in the BMN limit.      
This section also introduces necessary notation and presents single-double,
single-triple and double-triple trace BMN correlators. In section 3,
we discuss the scaling behavior of arbitrary multi-trace
correlators of BMN operators with $g_2$ and $J$. We demonstrate that
connected contributions to all of the correlators of this sort are
suppressed as $J\to\infty$. In section 4, we utilize this result to obtain the
anomalous dimension of an $i$-trace BMN operator at
$\O(g_2^2\lm')$. We also discuss some implications for the corresponding
processes in string theory in this section. Last section studies the duality between
$P^-$ and $\Delta-J$ at the level of matrix elements.    

Appendix A proves the scaling behaviour that we discuss in section 3.  
Appendix B deals with the basis transformation which
takes from the BMN basis into string basis in GT. Using the inputs
from \cite{PSVVV} and \cite{Gomis} we derive new decomposition
identities relating various multi-trace inner products with the
product of smaller order inner products. In particular, 
the free single-triple inner product decomposes into
single-double and double-triple inner products as,
$$G^{13}=\half G^{12}G^{23}.$$ Similarly we derive the identity,
$$G^{22}=\half(G^{21}G^{12}+G^{23}G^{32})$$ and discuss immediate
generalizations. We emphasize that these identities are derived by
relating the basis transformations proposed in \cite{Gomis} and
\cite{PSVVV}, therefore subject to explicit GT computations. These
computations which involve non-trivial summations are outlined in
Appendix C and these identities are proven there. In this appendix, 
we also explain evaluation of other sums that are used in sections
2,4 and 5. Appendix D computes $\O(g_2^2)$ and $\O(g_2^2\lm')$ contributions to
single-triple and $\O(g_2)$ and $\O(g_2\lm')$ contributions to
double-triple correlation functions.  

\section{Operator mixing at $g_2^2$ level}

In this section we shall carry out the diagonalization procedure of the multi-trace BMN operators 
including the mixing with triple trace operators. This is achieved by extending the method of 
\cite{CFHM} to include the $\O(g_2^2)$ and $\O(g_2^2\lm')$ effects in the diagonalization. In \cite{CFHM}, it was shown 
that the procedure of determining the eigenvectors and eigenvalues of the mixing matrix of single and 
double trace operators (which is $\O(g_2\lambda')$) is equivalent to first order non-degenerate 
perturbation theory. To include the mixing with triple trace operators one needs to go one step further 
in perturbation expansion, i.e. to second order perturbation theory. 

Let us first outline the method of \cite{CFHM} briefly. Consider the eigenvalue problem, 
\be\la{trns}
M_j^i e_{(k)}^j = \lambda_{(k)} e_{(k)}^i
\ee
where $M$ is the $3\times\infty$ dimensional mixing matrix of single,
double and triple trace operators.
\footnote{In the next section we explain why BPS type double and
  triple trace operators do not affect the following discussion.}
Here $i,j$ is a collective index labeling the state of a BMN operator, \eg for a triple trace, 
$i=\{m,y,z\}$ where $y=J_1/J$ and $z=J_2/J$ in (\ref{iBMN}) for
$i=3$. The order in $g_2$ of the various blocks of $M$ is indicated by,
\bd
M=\left(\begin{array}{ccc}
1 & g_2 & g_2^2 \\
g_2 & 1 & g_2 \\
g_2^2 & g_2 & 1
\end{array}\right).
\ed
Therefore it is possible to solve the eigenvalue problem order by order in $g_2$. Expanding $M$, $e$ and $\lm$ as
\bea
M^i_j & = &\rho_i\de^i_j+g_2 {M^{(1)}}^i_j + g_2^2 {M^{(2)}}^i_j\nonumber\\
e^i_{(k)} & = & \de^i_{k} + g_2 {e^{(1)}}^i_{(k)} + g_2^2 {e^{(2)}}^i_{(k)}\nonumber\\
\lm_{(k)} & = & \lm^{(0)}_{(k)} +g_2 \lm^{(1)}_{(k)} + g_2^2 \lm^{(2)}_{(k)},\nonumber
\eea
we obtain,
\bea\la{eiv2}
0 &=& \left(\rho_k-\lm^{(0)}_{(k)}\right)\de_k^i+g_2\left(\rho_i {e^{(1)}}^i_{(k)}
+{M^{(1)}}^i_k-\lm_{(k)}^{(0)}{e^{(1)}}^i_{(k)}-\de^i_{k}\lm_{(k)}^{(1)}\right)\nonumber\\
{}& &+g_2^2\left( \rho_i {e^{(2)}}^i_{(k)}+{M^{(2)}}^i_k+{M^{(1)}}^i_j{e^{(1)}}^j_{(k)}-
\lm_{(k)}^{(0)}{e^{(2)}}^i_{(k)}-\lm_{(k)}^{(1)}{e^{(1)}}^i_{(k)}-\lm_{(k)}^{(2)}\de^i_k\right).
\eea
At zeroth order one gets $\lm_{(k)}^{(0)}=\rho_k$. Using this in the
next order for $i\ne k$ yields the 
first order eigenvectors, 
$${e^{(1)}}^i_{(k)}=\frac{{M^{(1)}}^i_k}{\rho_k-\rho_i},$$
whereas for $i=k$ we learn that $\lm_{(k)}^{(1)}=0$. 
 
Using these results, $\O(g_2^2)$ piece of (\ref{eiv2}) for $i=k$ gives,
\be\la{seceiv}
\lm_{(k)}^{(2)}=\sum_j\frac{{M^{(1)}}^k_j{M^{(1)}}^j_k}{\rho_k-\rho_j},
\ee
and for $i\ne k$ we obtain the second order contribution to the eigenvectors,
\be\la{eic}
{e^{(2)}}^i_{(k)}=\frac{1}{\rho_k-\rho_i}\left({M^{(2)}}^i_k
+\sum_j\frac{{M^{(1)}}^i_j{M^{(1)}}^j_k}{\rho_k-\rho_j}\right).
\ee

Using above expressions for $e^{(1)}$ and $e^{(2)}$, we obtain the single-trace eigen-operator modified at $\O(g_2^2)$ as, 
\bea\la{modBMN1}
\tilde{O}^J_{n} &=& O^J_{n} + g_2 \sum_{my}\frac{\G^{my}_{n}}{n^2-(m/y)^2}O^J_{my}\nonumber\\
{}&&+g_2^2\sum_{myz}\frac{1}{n^2-(m/y)^2}\left(\G^{myz}_n +\sum_{m'y'}
\frac{\G^{myz}_{m'y'}\G^{m'y'}_n}{n^2-(m'/y')^2}\right)O^J_{myz}.
\eea
Here,
$$\G^i_j =G^{ik}\G_{kj}$$
are the matrix of anomalous dimensions where $G^{ij}$ denotes the
inverse metric on the field space. 
The metric, $G_{ij}$ is determined by the correlation functions $\<\bar{O}_i O_j\>$ at the free level whereas 
$\O(\lm')$ radiative corrections to this correlator yield
$\G_{ij}$. To wit,
\be\la{deff}
\<\bar{O}_i O_j\>=G_{ij}-\lm'\G_{ij}\ln(x^2\Lambda^2).
\ee
$G$ and $\G$ should be expanded in powers of $g_2$. Instead of
denoting the order in $g_2$ on $G$ and $\G$, we will show $g_2$ dependence explicitly in
what follows. 

Again, using above expressions for first and second order
eigenvectors, $e^{(1)}$ and $e^{(2)}$, one obtains the double-trace
eigen-operator as, 
\be\la{modBMN2}
\tilde{O}^J_{ny} = O^J_{ny} + g_2 \sum_{m}\frac{\G^{m}_{ny}}{(n/y)^2-m^2}O^J_{m}
+g_2\sum_{myz}\frac{\G^{myz}_{ny}}{(n/y)^2-(m/y')^2} O^J_{my'z'}.
\ee

A very important point is to notice that these expressions are valid
when the coefficients in front of $O_i$ on the RHS are finite for all
values of external and internal momenta. 
In particular one needs to check the finiteness of (\ref{modBMN1})   
when the incoming and outgoing world-sheet energies are equal, $n=\pm (m/y)$ and also at $n=\pm (m'/y')$ for the 
internal denominator in the third term. Note that, the danger of
degeneracy is absent only for the case $n=1$. Therefore without
checking the finiteness at $\O(g_2^2)$ one can assume the validity of
(\ref{modBMN1}) and (\ref{modBMN2}) only for the very particular case
of $n=1$! We will now demonstrate that the last term in (\ref{modBMN1})
is indeed divergent at the pole!\cite{German3}

Finiteness of the $\O(g_2)$ piece of (\ref{modBMN1}) 
was demonstrated in (\cite{CFHM}) where the coefficient was found to be, 
\be\la{coef12}
\frac{\G^{my}_{n}}{n^2-(m/y)^2} = -\frac{m/y}{n+(m/y)}G^{12}_{n;my}.
\ee
Here, $G_{n;my}$ denotes the tree-level inner product between single and double trace operators 
which was first computed in \cite{Constable}, 
\be\la{C12}
G^{12}_{n;my}=\frac{g_2}{\sqrt{J}}\sqrt{\frac{1-y}{y}}\frac{\sin^2(\pi n y)}{\pi^2(n-(m/y))^2}.
\ee 
We will also make abundant use of the radiative corrections to
single-double correlator which was also obtained in \cite{CFHM},
\be\la{G12}
\G^{12}_{n;my}=\left((\frac{m}{y})^2-n\frac{m}{y}+n^2\right)G^{12}_{n;my}.
\ee 

Let us now investigate $\O(g_2^2)$ part of (\ref{modBMN1}). First of all, it is 
not hard to show that there is no divergence at $n=\pm m'/y'$ in the second sum of the second term. 
These internal poles are canceled out by zeros of the
numerator. Similarly one can show that (\ref{modBMN1}) is finite at the
external pole $n=-m/y$. This is done at the end of this
section. However we shall shortly demonstrate that the external pole
at $n=+m/y$ give rise to a divergence \cite{German3} hence render the use of
non-degenerate perturbation theory invalid for $n>1$.\footnote{For
  $n=1$ it is impossible to satisfy the degeneracy condition $n=m/y$.} 

To go further we need (in addition to matrix elements already computed in the literature) 
the $O(g_2^2)$ contributions to  
$$G^{13}_{n;myz},\;\;\;{\mathrm{and}}\;\;\; \G^{13}_{n;myz}$$ 
and $O(g_2)$ contributions to
$$G^{23}_{ny;my'z'},\;\;\;{\mathrm{and}}\;\;\;\G^{23}_{ny;my'z'}.$$
Necessary computations are summarized in Appendix D and the results read, 
\bea
G^{13}_{n;myz}&=&\frac{g_2^2}{\pi^2J}\sqrt{\frac{z\tilde{z}}{y}}\frac{1}{(n-k)^2}
\bigg( (1-y)\sin^2(\pi n y)+y(\sin^2(\pi n z)+\sin^2(\pi n \tilde{z}))\nonumber\\
{}&&-\frac{1}{2\pi(n-k)}(\sin(2\pi n y)+\sin(2\pi n z)+\sin(2\pi n \tilde{z}))\bigg)\la{C13}\\
\G^{13}_{n;myz}&=&\lm'(n^2+k^2-nk)G^{13}_{n;myz}+B^{13}_{n;myz}\la{G13}.
\eea
Here $k=m/y$ is the world-sheet momentum of the double-string state and we defined 
$\tilde{z}=1-y-z$. 

Let us digress to underline an important detail. As we showed in
Appendix D, among the contributions to the radiative corrections 
to single-triple correlator there are contractible, semi-contractible
and non-contractible Feynman diagrams (see Appendix D for definition
of contractibility in planar diagrams). The contributions 
of the first two are summarized in the first term above, whereas 
$B^{13}_{n;myz}$ denotes the non-contractible contribution, 
\be\la{B13}
B^{13}_{n;myz}=\frac{2g_2^2\lm'}{\pi^3J}\frac{1}{(n-m/y)}\sqrt{\frac{z\tilde{z}}{y}}
\sin(\pi n z)\sin(\pi n \tilde{z})\sin(\pi n (1-y)).
\ee

Double-triple coefficients receive $\O(g_2)$ only from disconnected diagrams where the 2-3 process
is separated as 1-1 and 1-2. Therefore these require somewhat simpler computations and the 
results are,
\bea
G^{23}_{ny';myz}&=&{y'}^{3/2}G^{12}_{n;my/y'}(\de_{y',y+z}+\de_{y',1-z})
+\frac{g_2}{\sqrt{J}}\de_{mn}\de_{yy'}\sqrt{(1-y)z\tilde{z}}\la{C23}\\
\G^{23}_{ny';myz}&=&\frac{n^2}{{y'}^2}G^{23}_{ny';myz}+{y'}^{3/2}(\de_{y',y+z}+\de_{y',1-z})\frac{m}{y}(m/y-n/y')\la{C231}\\
{}&=&\left(\frac{n^2}{{y'}^2}-\frac{n}{y'}\frac{m}{y}+\frac{m^2}{{y}^2}\right)G^{23}_{ny';myz}.\la{G23}
\eea
 
We now move on to compute the $\O(g_2^2)$ term in
(\ref{modBMN1}). First of all one shows the curious
fact\footnote{which finds a natural explanation in the formulation of
  \cite{German3}} that 
\be\la{zeroine}
\G_n^{myz
}=0.
\ee
$\G^{myz}_n$ is decomposed as, 
\be\la{dec}
\G^{myz}_n=G^{myz;m'}\G_{m';n}+G^{myz;m'y'}\G_{m'y';n}+G^{myz;m'y'z'}\G_{m'y'z';n}.
\ee
One can easily invert $3\times\infty$ dimensional matrix $G_{ij}$, by solving the equation $G^{ik}G_{kj}=\de^i_j$ 
order by order in $g_2$. To $\O(g_2^2)$ one finds, 
\bd\la{invmat}
G^{ij}=\left(\begin{array}{ccc}
\de_{mn} + G^{12}_{m;py''}G^{12}_{py'';n} & -G^{12}_{m;ny'} & \half
G^{12}_{m;py''}G^{23}_{py'';ny'z'}-\half G^{13}_{m;ny'z'} \\
{}&{}&{}\\
-G^{12}_{my;n} & \de_{mn}\de_{y,y'}+G^{12}_{my;p}G^{12}_{p;ny'}+ &
-\half G^{23}_{my;ny'z'} \\
{} & +\half G^{23}_{my;py''z''}G^{23}_{py''z'';ny'} &{}\\
{}&{}&{}\\
\half G^{23}_{myz;py''}G^{12}_{py'';n}-\half G^{13}_{myz;n} & -\half
G^{23}_{myz;ny'} 
& \frac{1}{4}\de_{m,n}\de_{y,y'}(\de_{z,z'}+\de_{z,\tilde{z}'})+\\
{}&{}& +\frac{1}{4} G^{23}_{myz;py''}G^{23}_{py'';ny'z'} 
\end{array}\right).
\ed
By the use of decomposition identities listed in Appendix C, one can
prove that (\ref{dec}) vanishes (see App. C for details).          

Let us now consider the last term in (\ref{modBMN1}). A calculation
similar to the one that leads to (\ref{coef12}) gives,
\be\la{coef23}
\G^{myz}_{py'}=\half k'(k'-p)\frac{G^{12}_{p;my/y'}}{\sqrt{y'}}(\de_{y',y+z}+
\de_{y',y+\tilde{z}}),
\ee
where $k'=my'/y$. The other necessary ingredient, $\G^{py}_n$ was
already computed in \cite{CFHM}
$$ \G^{py'}_n= k(k-n) G_{n;py'}$$ where $k=p/y'$. Inserting these expressions into (\ref{modBMN1}), 
we get the whole coefficient in front of $O^J_{myz}$ as, 
\be\la{I2}
I=\frac{1}{n^2-(m/y)^2}\int_0^1 \ud y'\sum_{p=-\infty}^{\infty}\frac{k'(k'-p)k(k-n)}{2\sqrt{y'}(n^2-(p/y')^2)}
G_{p;my/y'}G_{n;py'}(\de_{y',y+z}+ \de_{y',y+\tilde{z}}).
\ee
Despite the appearance of $n^2-(p/y')^2$ in the denominator there is no divergence at $n=\pm p/y'$ because 
$G_{n;py'}$ in the numerator also vanishes at these intermediate
poles. We will now show however that {\emph{$I$ is divergent
at $n=m/y$}\cite{German3}.
The residue of $I$ at $n=m/y$ is, 
\bea
 (n-m/y)I\bigg|_{n=m/y}&=& -\frac{g_2^2}{4\pi^4J}\sqrt{\frac{z\tilde{z}}{y}}\int_0^1\ud y' \frac{1}{y'}
(\de_{y',y+z}+ \de_{y',y+\tilde{z}})\sin^2(\pi n y')\nonumber\\
{}&& \sum_{p=-\infty}^{\infty}
\frac{p}{y'}\frac{\sin^2(\pi p y/y')}{(n^2-(p/y')^2)(n-p/y')^2}\nonumber.
\eea

We emphasize that the infinite series in this expression is a prototype for the non-trivial sums 
that appear in the computations involving triple-trace BMN operators. We explain the computation of 
this one and other similar sums which will be necessary for the next section in Appendix B.
\footnote{Unfortunately none of the well-known symbolic computation programs is helpful.} 
The result is,
\be\la{sum11} 
\sum_{p=-\infty}^{\infty}
\frac{p}{y'}\frac{\sin^2(\pi p y/y')}{(n^2-(p/y')^2)(n-p/y')^2}=y'
\left(\frac{\pi^3y^2}{2}\cot(\pi n y')-\frac{\pi^2 y}{4n}\right).
\ee  
Inserting this in the above expression for the residue and evaluating the integral gives, 
$$(n-m/y)I\bigg|_{n=m/y}=\frac{g_2^2}{J}\frac{y}{8n\pi^2}
\sqrt{\frac{z\tilde{z}}{y}}\sin(\pi n z)\sin(\pi n \tilde{z}).$$ 
Since the residue does not vanish at $n=+m/y$, (\ref{modBMN1}) becomes
divergent at this pole. 

To see that there is no further divergence in (\ref{modBMN1})
let us consider what happens at the other pole
$n=-m/y$. It is easy to see from (\ref{coef12}) that the second term
is finite because $G^{12}$ in the numerator linearly goes to zero as
well as the denominator. The complicated second piece in the last term of
(\ref{modBMN1}) seems to be divergent at the first sight. Let us look
at the residue at this pole,
\bea
 (n+m/y)I_2\bigg|_{n=-m/y}&=& -\frac{g_2^2}{4\pi^4J}\sqrt{\frac{z\tilde{z}}{y}}\int_0^1\ud y' \frac{1}{y'}
(\de_{y',y+z}+ \de_{y',y+\tilde{z}})\sin^2(\pi m y'/y)\nonumber\\
{}&& \times\sum_{p=-\infty}^{\infty}
\frac{p}{y'}\frac{\sin^2(\pi p y/y')}{((m/y)^2-(p/y')^2)^2}\nonumber.
\eea 
This sum vanishes thanks to the antisymmetry of the summand. Thus we
saw that all of the terms in (\ref{modBMN1}) is finite at the
pole $n=-m/y$\footnote{Of course one has to worry about finiteness of (\ref{modBMN2}) at $n=\pm m/y$.
But this requires much little effort to see from the expressions for $\G^{23}$ and $\G^{12}$.}. 

The fact that $I=\infty$ at $n=m/y$, hence (\ref{modBMN1}) is
ill-defined at this pole hints that \emph{one should rather use degenerate
perturbation theory to handle the diagonalization problem}
\cite{German3}. Although somewhat disappointing, this result is by no means unexpected. On the
GT side one can reason as follows.\footnote{This is a suggestive argument due to Dan Freedman.}  
't Hooft limit suggests that anomalous dimensions of observables be
expanded in powers of $g_2^2$ where $g_2=1/N$ in 't Hooft limit and
$g_2=J^2/N$ in the BMN double scaling limit. Had single-trace BMN
operators been degenerate with double-trace BMN operators one would 
expect an $\O(g_2)$ shift
in the single-trace anomalous dimension. This would be unexpected for the 't Hooft
expansion of the observables hence might have indicated an inconsistency
in the BMN theory. However the degeneracy of sigle and triple-trace
BMN operators, at most, gives rise to an $\O(g_2^2)$ shift in
$\Delta_1$ which is not inadmissible. By the same token, one generally
expects degeneracy among BMN operators with only odd-numbered traces
and only even-numbered traces seperately. At order $g_2^2$, this result indeed follows
by the scaling law of GT correlators derived in the next section
provided that there is no degeneracy between single and double-traces
and there is degeneracy between single and triple traces. Hence, at
this order the degenerate subspace of BMN operators are divided into
two subspaces which include odd and even numbered traces seperately. 

On the string theory side the degeneracy of single and triple-trace
operators indicate that a single-string state can decay into a
triple-string state the same world-sheet momentum $m=ny$. Furthermore,
as we mentioned in the introduction, correspondence of trace number in
GT and string number in ST loses its meaning for finite
$g_2$. Therefore the general conclusion is that 
an initial string state that is composed of states of
different string number but all on the same ``momenta-shell'',
$n=m/y$, 
is generically unstable and can decay into states that are stable at
$O(g_s^2)$. 
These stable states should be in correspondence with the eigen-operators of the
degenerate subspaces in GT side. This conclusion is hardly surprising. 

We shall not pursue this degeneracy problem further in this paper, but
based on some preliminary calculations we make the following
conjecture. Consider the degeneracy problem for finite $J$ (which we
choose as an odd number for convenience). Then two degenerate
subspaces involve 1,3,\dots,$J$-trace and
2,4,\dots,$J$-1-trace operators seperately. To find whether degeneracy
gives rise to a shift in the eigenvalues one should
diagonalize the order $g_2^2$ ``transition matrix'',
$M_i^{i+2}$ (finite $J$ version of
(\ref{trns})) at $n=m/y$ seperately for odd and even $i$. We conjecture that
regardless the exact form of $M$, there exist an eigenstate $O'_1$ that
tend to the BMN operator $\tilde{O}_1$ (at order $g_2^2$) as
$J\to\infty$ 
where $\tilde{O}_i$ is the mostly $i$-trace eigen-operator of the
dilatation generator that is obtained by the non-degenerate
formulation at $\O(g_2)$. Futhermore there exist $i$-trace
eigenstates, $O'_i$ in the degenerate subspace whose eigenvalues 
tend to the anomalous dimensions of $\tilde{O}_i$ that are obtained by
naively using the non-degenerate formulation. 
Therefore the eigenvalues of these particular $O'_i$ will coincide
with the anomalous dimensions which can be
obtained via non-degenerate theory ignoring the aforementioned
mixing. For the case of $i=1$ this can be understood as a
justification of (\ref{contour}). For $i>1$ this leads to a
simplification in determination of the higher-trace anomalous
dimensions which we employ in section 4.
We prefer to leave this assertion as a conjecture in this paper.    
         
In section 4 we will use this conjecture to make predictions about
string-theory amplitudes. We will first compute the anomalous
dimension of a general $i$-trace BMN operator, $\Delta_{i}-J$ by the
method of non-degenerate perturbation theory. Since
dilatation generator is supposed to correspond to $P^-$, this will
give a prediction for the eigenvalue of $P^-$ in the two-string
sector. Next, we will move on to compute the modified mixing matrices,
$\tilde{\G}^{22}$ and $\tilde{\G}^{13}$ in the string basis. This will
allow us to make predictions for the corresponding matrix elements of
$P^-$.  

\section{Dependence on $g_2$ and $J$ of an arbitrary gauge theory correlator} 
  
As shown by detailed investigation in 
the recent literature n-point functions of the observables in the BMN limit come with 
definite dependence on the dimensionless parameters, $\lm'$ and $g_2$.
\footnote{For finite $g_2$ proof exist only at linear order in $\lm'$ although it is very 
likely to hold at higher loops. For $g_2=0$, \cite{Zanon} showed that
sum of radiative corrections to single-single BMN correlator at all
orders in $\O(g_{YM}^2)$ can be expressed as a function only $\lm'$.} 
Generally, the correlators also have an explicit dependence on $J$
which will turn out to be crucial in drawing conclusions about the
corresponding string processes. 

In this section we will discuss the $g_2$ and $J$ dependence of a
two-point correlator of multi-trace BMN operators with two scalar impurities:
\be\la{corr}
C^{ij}\equiv\<:\bar{O}_n^{J_1}\bar{O}^{J_2}\cdots\bar{O}^{J_i}(x):
:O_m^{J_{i+1}}O^{J_{i+2}}\cdots O^{J_{i+j}}(0):\>\bigg|_{connected}.
\ee
This scaling law that we find is also valid 
for the correlators of a more general class of operators, 
$$:O_{n_1}^{J_1}\dots O_{n_i}^{J_i}O^{J_{i+1}}_{\phi}\cdots O^{J_{i+j}}_{\phi}O^{J_{i+j+1}}_{\psi}
\cdots O^{J_{i+j+k}}_{\psi}O^{J_{i+j+k+1}}\cdots O^{J_{i+j+k+l}}:,$$
for arbitrary $i,j,k,l$ and also for the n-point functions
involving same type of operators. This should be clear from the discussion in Appendix A.  

The space-time dependence of (\ref{corr}) is trivial: $(4\pi^2x^2)^{-J-2}$ 
in free theory and $(4\pi^2x^2)^{-J-2}\ln(x^2\Lambda^2)/(8\pi^2)$ 
at $\O(\lm')$ where $J$ is the total number of $Z$ fields, \ie
$J=J_1+\cdots+{J_i}= J_{i+1}+\cdots+J_{i+j}$ in (\ref{corr}). 
Without loss of
generality, one can assume $j\le i$. 
There are various connected and disconnected diagrams with 
different topology that contribute to (\ref{corr}). 
Since the results for disconnected contributions will be given by
(\ref{corr}) for smaller $i$ and $j$, it suffices 
to consider the fully-connected contribution to 
(\ref{corr}). In Appendix A we prove that the fully-connected piece of
(\ref{corr}) has the following general form,
\be\la{general}
C^{ij}=\frac{g_2^{i+j-2}}{J^{(i+j)/2-1}}\left\{G^{ij}\frac{1}{(4\pi^2x^2)^{J+2}}
-\lm'\G^{ij}\frac{1}{(4\pi^2x^2)^{J+2}}\ln(x^2\Lambda^2)\right\}.
\ee
Here the ``free'' and ``anomalous'' matrix elements, $G$ and $\G$ are
functions of world-sheet momenta, $m$, $n$ and of the ratios $J_s/J$
for $s=1,\dots i+j$. Disconnected pieces are less suppressed by $J$. 
Note that {\emph{suppression of $C^{ij}$ in the
    BMN limit is absent only when $i=j=1$.}} We state the
conclusion as,

\begin{itemize}
\item{\emph{Connected contributions to $C^{ij}$ are suppressed in the
      BMN limit for $i$ and/or $j$ larger than 1.}}
\end{itemize}

Let us briefly discuss the case of BPS type multi-trace operators. A
BPS type multi-trace operator which involve two scalar type impurities
is defined as, 
\be\la{iBPS}
O_{i;\phi\psi}^J=:O_{\phi}^{J_1}O_{\psi}^{J_2}O^{J_3}\cdots O^{J_i}:\;,
\ee
where, 
\be\la{BPSop}
O_{\phi}^J=\frac{1}{\sqrt{N^{J+1}}}\Tr(\phi Z^J)
\ee
and a similar definition for $O_{\psi}^J$. 

One observes that a generic matrix element between BMN and BPS
$i$-trace operators (both having the same number of traces) at all
orders in $g_2$ are suppressed by a power of $J$. This can be seen
by noting that these correlators should necessarily be partially
connected (since both $O^{J'_1}_{\psi}$ and $O^{J'_2}_{\phi}$ in the BPS type 
$i$-trace operator should connect to the same $O_n^{J_1}$ in the
$i$-trace BMN operator) 
and therefore suppressed by at least a factor of $J$ with respect to BPS-BPS or
BMN-BMN correlators of the same number of traces. Above scaling law
tells us that, in the latter cases 
suppression at an arbitrary order in $g_2$ can only be avoided by
completely disconnected graphs with an arbitrary number of loops. 
This simple observation allowed us to ignore BPS type double and
triple operators in the previous section that would otherwise
contribute in the intermediate sums. 

\section{Anomalous dimension of a general multi-trace BMN operator at $\O(g_2^2)$}

The fact that disconnected contributions to the multi-trace
correlators are suppressed as $J\to\infty$ has direct consequences for
the scale dimension of multi-trace operators both at order $g_2^2$ and
higher. 

Call the $i$-trace BMN operator in (\ref{corr}) as $O_i$. 
Here $i$ is a collection of
labels, $i=\{n,y_1,\dots,y_{i}\}$ with $y_1\equiv\frac{J_1}{J}$, etc. Because of the
non-vanishing mixing, $\<\bar{O}_iO_j\>$ with multi-trace
operators of different order ($i\ne j$), $O_i$ is {\emph{not}} an
eigen-operator of $\Delta-J$ and a non-trivial diagonalization
procedure is required to obtain the true scale
dimension. Eigenvectors at $\og2$ is affected only by mixing of $O_i$
with $O_{i\pm 1}$. The diagonalization procedure is essentially
equivalent\footnote{see the discussion at the end of section 2 for the
  effects of mixing with higher trace BMN operators}
to non-degenerate perturbation theory \cite{CFHM} and as in
section 2 one obtains the mostly $i$-trace eigen-operator as, 
\be\la{eiop}
\tilde{O}_i=O_i+g_2\sum_{j=i\pm 1}\frac{\G^j_i}{\rho_i-\rho_j}O_j,
\ee
where $\G^j_i=G^{jk}\G_{ki}$ and $\rho_k$ is the $O(g_2^0)$
 eigenvalue of the $k$-trace operator ($\rho_i=(n/y_1)^2$ in case of
 $O_i$ in (\ref{corr}))\footnote{The proof of the validity of
 non-degenerate perturbation theory at $\O(g_2^2)$ is illustrated in
 case of single-trace and double-trace operators in section 2. However
 this proof immediately generalizes to the general case of $i$-trace
 operators because of the suppression of connected
 correlators. Requirement of disconnectedness boils down the required 
 computation to the one presented in section 2.}. To compute the
 eigenvalue we need, (using (\ref{eiop})),
\be\la{mixedcorr}
\<\bar{\tilde{O}}_i\tilde{O}_i\>=\<\bar{O}_iO_i\>
+2g_2\sum_{j=i\pm 1}\frac{\G^j_i}{\rho_i-\rho_j}\<\bar{O}_jO_i\>
+g_2^2\sum_{j,k=i\pm1}\frac{\G^j_i}{\rho_i-\rho_j}\frac{\G^k_i}{\rho_i-\rho_k}\<\bar{O}_jO_k\>. 
\ee
Call the $\O(g_2^2)$ part of this quantity as,
\be\la{mixedcorr2}
\<\bar{\tilde{O}}_i\tilde{O}_i\>\bigg|_{g_2^2}
=\tilde{G}_{ii}-\lm'\tilde{\G}_{ii}\ln(x^2\Lambda^2).
\ee
Generalizing the method described in \cite{CFHM} to the case of
multi-trace operators, we express the true scale
dimension at this order in terms of the above quantities as, 
\be\la{eivv}
\Delta_{i}\bigg|_{g_2^2}=\tilde{\G}_{ii}-\rho_i\tilde{G}_{ii}.
\ee
It is not hard to see that (\ref{eivv}) is equivalent to (\ref{seceiv}) as
it should be. To compute $\Delta_i$ from either (\ref{eivv}) or
(\ref{seceiv}) one needs $G_{ij}$ and $\G_{ij}$ to the necessary order. 
Although second method is a short-cut we prefer to start from
(\ref{eivv}) because this method makes it clear that modified operators,
$\tilde{O}_i$ are true eigen-operators of the dilatation generator.
   
Now, the crucial point is to recall that the connected contributions
to $\tilde{\G}_{ii}$ and $\tilde{G}_{ii}$ are suppressed as
$J\to\infty$ and the evaluation of these quantities reduce to the
evaluation of only the fully disconnected pieces. 
For example the quantity $G_{ii}$
receives non-zero contributions only from the following completely
disconnected Wick contractions,
\bea
G_{ii}\bigg|_{g_2^2}&=&\<\bar{O}_iO_i\>_{g_2^2}
=\<\bar{O}^{J_1}_nO^{J_1}_n\>_{g_2^2}\<\bar{O}^{J_2}{O}^{J_2}\>_{g_2^0}
\cdots\<\bar{O}^{J_i}O^{J_i}\>_{g_2^0}\nonumber\\
{}&&+\<\bar{O}^{J_1}_nO^{J_1}_n\>_{g_2^0}\<\bar{O}^{J_2}{O}^{J_2}\>_{g_2^2}
\cdots\<\bar{O}^{J_i}O^{J_i}\>_{g_2^0}+\cdots\nonumber\\
{}&&+\<\bar{O}^{J_1}_nO^{J_1}_n\>_{g_2^0}\<\bar{O}^{J_2}{O}^{J_2}\>_{g_2^0}
\cdots\<\bar{O}^{J_i}O^{J_i}\>_{g_2^2}.
\la{pattal}
\eea

\begin{figure}[htb]
\centerline{\epsfxsize=15cm\epsfysize=6cm\epsffile{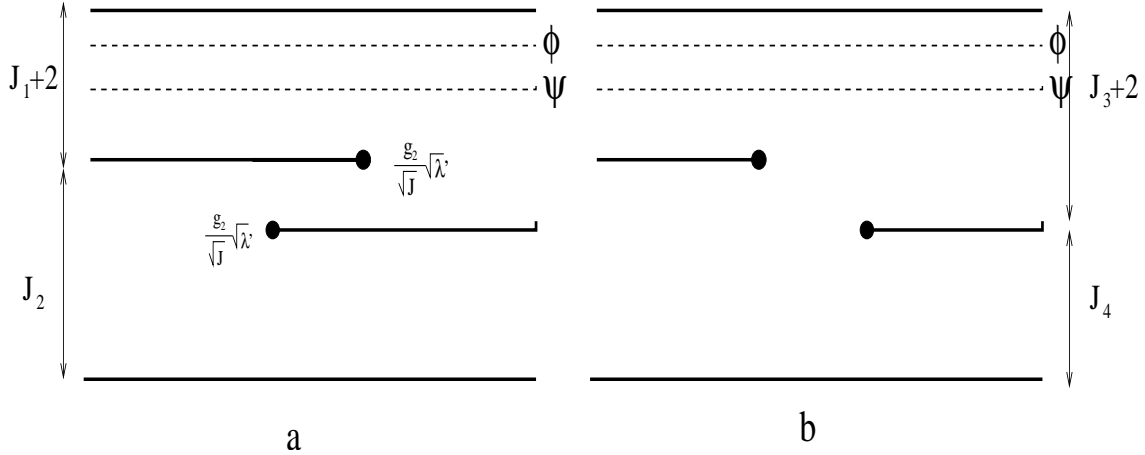}}
\caption{Left figure shows that connected contribution to $2\to
  3\to 2$ process is suppressed by $1/J$. Right figure shows similar
  suppression of mixing of double trace operators with single-traces.} 
         \label{FIG4}
\end{figure}

Now, we shall compute $\Delta_{i}$ at $\O(g_2^2)$ for arbitrary
$i$. One first observes that $j=i-1$ channel in (\ref{mixedcorr})
necessarily gives connected diagrams hence suppressed by the power of
$J$ given by (\ref{general}). More explicitly, the summands in the
$(i-1)$ channel are $\O(g_2^2/J)$ but the intermediate sums do not
provide a compensating factor of $J$ unlike in the $(i+1)$
channel. This is illustrated in Fig.2 in case of $i=2$\footnote{For an
  explanation for these ``string-like'' transition diagrams see the end of this
  section}. 
Similarly the summands in $(i+1)$ channel are also $\O(g_2^2/J)$, therefore only
disconnected $i\to (i+1)\to i$ processes can contribute. This is also 
illustrated in Fig.2. The conclusion
is that,
\begin{enumerate}
\item
 $i$-$(i-1)$ mixing does not affect $i$-trace eigenvalue,
\item Only disconnected $i\to (i+1)\to i$ processes in $i$-$(i+1)$
  mixing matter. (see Fig.2)
\end{enumerate}   

One obtains the quantities, $\tilde{G}_{ii}$
and $\tilde{\G}_{ii}$ that are necessary to evaluate (\ref{eivv}) from
(\ref{mixedcorr}). The former reads,
\bea
\tilde{G}_{ii}&=&G_{ii}+ 2\sum_{i+1}\frac{\G^{i+1}_i}{\rho_i-\rho_{i+1}}G_{i+1,i}
+\sum_{i+1,i'+1}\frac{\G^{i+1}_i}{\rho_i-\rho_{i+1}}\frac{\G^{i'+1}_i}
{\rho_i-\rho_{i'+1}}G_{i+1,i'+1}\nonumber\\ 
&=& G_{ii}+\sum_{i+1} \frac{\G^{i+1}_i}{\rho_i-\rho_{i+1}}
\left(2G_{i+1,i}+i!\frac{\G^{i+1}_i}{\rho_i-\rho_{i+1}}\right).
\la{bongo}
\eea

Here, we use the indices in a schematic sense, for example $(i'+1)$
and $(i+1)$ are independent indices that both refer to a collective index which labels an $(i+1)$-trace
operator, \ie $i+1=\{m,y_1,\dots,y_i\}$ and
$i'+1=\{m',y'_1,\dots,y'_i\}$. In the second line above, we used the
expression for the lowest order, free two-point function of
$(i+1)$-trace BMN operators. For general $i$, this is easily obtained by
recalling the fact that only disconnected pieces contribute. Thus to
lowest order, $\O(g_2^0)$, 
$G_{ii}$ is product of its disconnected pieces summed over all ways of
Wick contracting various BPS operators:
\be\la{Cii}
G_{ii'}=G_{my_1\cdots y_i;m'y'_1\cdots
  y'_i}=\de_{mm'}\de_{y_1y'_1}\sum_P\de_{y_2y'_{P(2)}}\cdots\de_{y_iy'_{P(i)}}, 
\ee
where $P$ runs over all permutations of the set $\{2,\dots i\}$. We
stress that we need the $\O(g_2^2)$ expression for $G_{ii}$ in
(\ref{bongo}) rather than (\ref{Cii}). Using this
formula for $G_{i+1,i'+1}$ in the first line of (\ref{bongo}) and summing over the
indices $i'+1$ produces a factor of $i!$. To go further we need 
$$ \frac{\G^{i+1}_i}{\rho_i-\rho_{i+1}}.$$ 
The only $\og2$ contributions to the matrix element come from the following two terms, 
$$\G^{i+1}_i=G^{i+1,j+1}\G_{j+1,i}+G^{i+1,j}\G_{j,i}.$$     
We need to invert the $2\times\infty$ dimensional matrix of inner
products between $i$-trace and $(i+1)$-trace operators.
\footnote{See section 3 for a justification of our omitting 
BPS type $i$-trace and $(i+1)$-trace operators in the evaluation of the eigenvalue.}
It is not hard to find the inverse perturbatively at $\og2$ with the result,
\bd
G^{A,B}=\left(\begin{array}{ccc}
\frac{G_{i,i'}}{(i-1)!(i-1)!}+\O(g_2^2) &
-\frac{G_{i,i'+1}}{(i-1)!i!}+\O(g_2^3) \\
{}&{}&{}\\
-\frac{G_{i+1,i'}}{(i-1)!i!}+\O(g_2^3) & \frac{G_{i+1,i'+1}}{i!i!}+\O(g_2^2) 
\end{array}\right).
\ed

We also need {\emph{free}} $i$-$(i+1)$ correlator at $\og2$. Since it is given
by the fully-disconnected contribution, it is obtained as a
simple generalization of (\ref{C23}): 
\bea
G^{i,i+1}_{my_1\cdots y_i;m'y'_1\cdots
  y'_{i+1}}&=&\de_{m,m'}\de_{y_1,y'_1}\frac{g_2}{\sqrt{J}}
\sum_{P,P'}\de_{y_{P(2)},y'_{P'(2)}}\cdots\de_{y_{P(i-1)},y'_{P'(i-1)}}
\de_{y_{P(i)},y'_{P'(i)}+y'_{P'(i+1)}}\nonumber\\
{}&&\times\sqrt{(1-y_{P(i)})y'_{P'(i)}y'_{P'(i+1)}}\nonumber\\
{}&+&y_1^{3/2}G^{12}_{m,m'y'_1/y_1}\sum_P\de_{y_2,y'_{P(2)}}
\cdots\de_{y_{i},y'_{P(i)}}\de_{y_1,y'_1+y'_{P(i+1)}}.
\la{Ciii}
\eea
Here the first term is a generalization of the second term in
(\ref{C23}) and the second terms is the generalization of the first
term in (\ref{C23}). The sum $P$ in the first term is over cyclic
permutations of the set $\{2,\dots,i\}$ \ie it has dimension $i-1$ and sum $P'$ is over all
possible ways of choosing two indices out of the set $1,\dots i+1$ (to
form the single-double BPS correlator with $P(i)$th BPS operator in
$O_i$) and than taking all possible permutations in the rest of the
indices, \ie $dim(P')=(i-2)!i(i-1)/2$.   

Finally we need the first order {\emph{radiative corrections}} to this
correlator. Much as in (\ref{G23}) this is given as, 
\bea
\G^{i,i+1}_{my_1\cdots y_i;m'y'_1\cdots
  y'_{i+1}}&=&\frac{m^2}{y_1^2}G^{i,i+1}_{my_1\cdots y_i;m'y'_1\cdots
  y'_{i+1}}+y_1^{3/2}G^{12}_{m,m'y'_1/y_1}\frac{{m'}}{{y'}_1}\left(\frac{{m'}}{{y'}_1}-\frac{{m}}{{y}_1}\right)\nonumber\\
{}&&\times\left\{\sum_P\de_{y_2,y'_{P(2)}}
\cdots\de_{y_{i},y'_{P(i)}}\de_{y_1,y'_1+y'_{P(i+1)}}\right\}\nonumber\\
{}&=&\left(\frac{m^2}{y_1^2}-\frac{mm'}{y_1y'_1}+\frac{{m'}^2}{{y'}_1^2}\right)G^{i,i+1}_{my_1\cdots
    y_i;m'y'_1\cdots y'_{i+1}}
\eea
\la{Giii}

Eqs. (\ref{Ciii}) and (\ref{Giii}) are sufficient to determine
$\G^{i+1}_i$ at $\og2$:
\bea
\G^{i+1}_i&=&G^{i+1,j+1}\G_{j+1,i}+G^{i+1,j}\G_{j,i}\nonumber\\
{}&=&\frac{1}{i!}y_1^{3/2}G^{12}_{m,m'y'_1/y_1}\frac{m'}{y'_1}(\frac{m'}{y'_1}-\frac{m}{y_1})
\left\{\sum_P\de_{y_{2},y'_{P(2)}}\cdots\de_{y_{i},y'_{P(i)}}\de_{y_1,y'_1+y'_{P(i+1)}}\right\}.
\eea

We insert this expression in (\ref{bongo}) and perform the sum over
the intermediate $(i+1)$ channel. Most of the terms in the contraction
of delta-functions will be suppressed (\eg first term in (\ref{Ciii})
multiplied with $\G^{i+1}_i$ and summed over $(i+1)$ is suppressed by
$1/J$). Result is, 
\be\la{Ciii2}
\tilde{G}_{ii'}=G_{ii'}-\delta_{ii'}\sum_{ny}y^3G^{12}_{m,ny/y_1}G^{12}_{m',ny/y'_1}
\frac{(\frac{n}{y})(\frac{n}{y}+\frac{m}{y_1}+\frac{m'}{y'_1})}{(\frac{n}{y}+\frac{m}{y_1})
(\frac{n}{y}+\frac{m'}{y'_1})}.
\ee
Here, $\delta_{ii'}$ is a shorthand for the delta functions that arise
from disconnected Wick contractions:
$$\delta_{ii'}=\de_{y_1,y'_1}\sum_P\de_{y_2,y'_{P(2)}}\cdots\de_{y_{i},y'_{P(i)}}.$$

A completely analogous computation yields $\tilde{\G}_{ii'}$ as,
\be\la{Giii2}
\tilde{\G}_{ii'}=\G_{ii'}-\delta_{ii'}\sum_{ny}y^3G^{12}_{m,ny/y_1}G^{12}_{m',ny/y'_1}
\frac{(\frac{n}{y})(\frac{n^3}{y^3}+\frac{m^3}{y_1^3}
+\frac{{m'}^3}{{y'}_1^3})}{(\frac{n}{y}+\frac{m}{y_1})
(\frac{n}{y}+\frac{m'}{y'_1})}.
\ee
For the same reason as above we only need disconnected contributions
to $G_{ii}$ and $\G_{ii}$ which are trivial to evaluate. In case of
$i=2$ required diagrams are illustrated in Fig.2.b,c and d. In terms of
the known expression for single string correlator at $\O(g_2^2)$ and
radiative corrections to this \cite{Constable}\cite{German1}, one
readily gets (see (\ref{pattal})) 
\be\la{Cii2}
G_{ii'}=g_2^2\delta_{ii'}\left(y_1^4A_{mm'}+\frac{\de_{mm'}}{24}\sum_{r=2}^iy_r^4\right),
\ee
and
\be\la{Gii2}
\G_{ii'}=g_2^2\delta_{ii'}\left(y_1^2(m^2-mm'+{m'}^2)
A_{mm'}+y_1^2 B_{mm'}
+\frac{\de_{mm'}}{24}\frac{m^2}{y_1^2}\sum_{r=2}^iy_r^4\right),
\ee
where $A$ and $B$ matrices are first defined in \cite{Constable} and
are reproduced in eqs. (\ref{AA}) and (\ref{BB}). 

Let us digress for a moment to discuss a simpler type of degeneracy in
energy eigenvalues that is referred as {\emph{momenta-mixing}}.
So far, we formulated our discussion in terms of the multi-trace BMN
operators as given in eqs. (\ref{iBMN}) and (\ref{BMNop}). In doing so
we ignored a degeneracy in the energy eigenvalues corresponding to
operators with opposite world-sheet momenta, namely $O^J_n$ and
$O^J_{-n}$ carry the same energy that is $n^2$. To incorporate the
effects of this momenta-mixing one should disentangle the degenerate
states by going to $\pm$ basis, 
\be\la{pm}
O^{\pm\, J}_n=\frac{1}{\sqrt{2}}(O^J_n\pm O^J_{-n}).
\ee
In $\pm$ basis BMN operators with two scalar impurities transform in
the singlet and triplet representation under the $SU(2)$ subgroup of
the full $SO(4)$ R-symmetry. One can easily reformulate our results in
this basis. For example the eigenvalue equation reads,    
\be\la{eiv}
\Delta^{\pm}_{my_1\dots y_i\; ;\;my_1\dots y_i}\bigg|_{g_2^2}=
\Delta_{my_1\dots y_i\; ;\;my_1\dots y_i}\bigg|_{g_2^2}\pm
\Delta_{my_1\dots y_i\; ;\;-my_1\dots y_i}\bigg|_{g_2^2}
\ee

Using eqs. (\ref{Cii2}), (\ref{Gii2}),(\ref{Ciii2}) and (\ref{Giii2}), 
it is straightforward to see that, 
\be\la{deg}
\Delta_{my_1\dots y_i\; ;\;-my_1\dots y_i}\bigg|_{g_2^2}=0.
\ee
Thus, our first observation is that 
\begin{itemize}
\item {\emph{Scaling dimension of all multi-trace BMN
operators remain degenerate in $\pm$ channels at $\O(g_2^2)$.}}
\end{itemize}
This fact can be explained by relating the multi-trace BMN operators
in $+$ and $-$ channels by a sequence of supersymmetry 
transformations\cite{vector}\cite{CFHM}\cite{Beisert}. However, we
would like to emphasize that this explanation holds only in the strict
BMN limit where $J\to\infty$. The reason is that, although
supersymmetry is exact for any $J$, BMN operators do not exactly transform under
long multiplets unless $J$ is strictly taken to $\infty$. We would
also like to emphasize that degeneracy of multi-trace BMN operators
can be viewed as a consistency check on our long computation because 
our results are valid also for single-trace BMN operators for $i=1$,
where this degeneracy is well-established \cite{German2}\cite{CFHM}. 

Having established the degeneracy in $\pm$ basis, we can compute the
eigenvalue by using (\ref{Cii2}), (\ref{Gii2}),(\ref{Ciii2}) and 
(\ref{Giii2}) in 
$$\Delta^{\pm}_{i}\bigg|_{g_2^2}=\tilde{\G}_{ii}-\rho_i\tilde{G}_{ii}.$$
Straightforward computation gives,
$$
\Delta^{\pm}_{ii}=y_1^2\left(g_2^2 B_{mm}
-J\int_0^1\ud x\sum_{n=-\infty}^{\infty}(G^{12}_{m,nx})^2\frac{k^2(k^2-m^2)}{(k+m)^2}\right),
$$
where $k=m/x$. Using (\ref{C12}) one gets,
\be 
\Delta^{\pm}_{i}\bigg|_{g_2^2}=
\frac{y_1^2}{4\pi^2}\left(\frac{1}{12}+\frac{35}{32\pi^2m^2}\right).
\ee
This is exactly the single-trace anomalous dimension that were
computed in \cite{German2} and \cite{CFHM} up to the normalization
factor $y_1^2$. This is hardly surprising given the fact that all
Feynman diagrams that contribute to the evaluation of $\Delta_{i}$
separate into completely disconnected pieces. Since the only piece
that can contribute to anomalous dimension is coming from the
single-trace BMN sub-correlator (BPS sub-correlators are protected), 
we obtain the single-trace anomalous dimension as a result. 
However, from a general point of view this is a striking result and is 
one of the main conclusions of this paper:
\begin{itemize}
\item {\emph{Scaling dimension of all multi-trace BMN operators are
      determined by the dimension of the single-trace operator as}}\\
$$\Delta^{\pm}_{i}=\left(\frac{J_1}{J}\right)^2\Delta_{11}.$$
\end{itemize}

We believe that this result will also hold at higher orders in $g_2$
because the fact that only disconnected Feynman diagrams survive the
BMN limit is still valid for higher orders in genus expansion. We see
this by noticing that each $g_2$ comes along with a factor of $1/\sqrt{J}$ in the
expansion, (see eq. (\ref{general}), also Fig.2 below). This should become more clear in
the following. 

This result establishes a firm prediction for $\O(g_2^2)$ eigenvalues
of the light-cone SFT Hamiltonian. When translated into string
language, this prediction reads,
$$\<\psi_i|P^-|\psi_i\>=\left(\frac{p^+_1}{p^+}\right)^2\<\psi_1|P^-|\psi_1\>.$$  
We would like to emphasize that this prediction is completely
independent of the field theory basis which identifies operators that are
dual to the string states. 

\begin{figure}[htb]
\centerline{ \epsfysize=8cm\epsffile{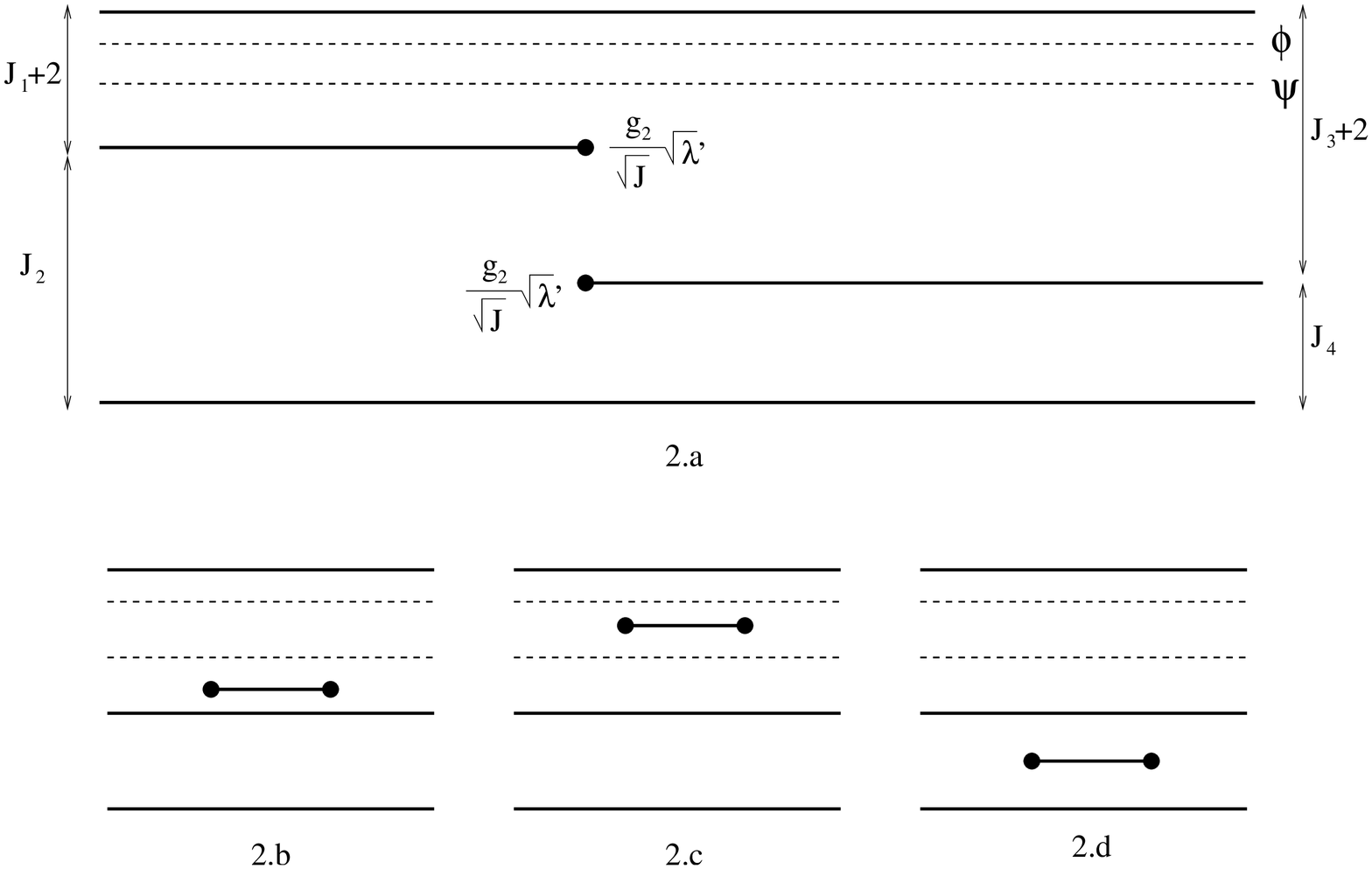}}
\caption{A representation of planar contributions to 
  $\<:\bar{O}^{J_1}_n\bar{O}^{J_2}::O^{J_3}_m O^{J_4}:\>$ at
  $\O(g_2^2\lm')$. Dashed lines represent scalar impurities. We do not
  show $Z$ lines explicitly. Vertices are of order
  $g_2\sqrt{\lm'/J}$. {\bf a} Connected contribution. {\bf b, c, d}
  Various disconnected contributions.}   
         \label{FIG1}
\end{figure}

Let us now discuss the implications of our findings for some of the string
amplitudes. For this let us represent our discussion about the
scaling of correlators with $g_2$ and $J$ in a diagrammatic
way that is suggestive for light-cone SFT. For instance we represent
the double-trace correlator in the BMN basis,
$\<\bar{O}_2O_2\>_{g_2^2\lm'}$, as in Fig.2 where Fig.2.a shows the
connected contribution to this correlator while Fig.2.b,c d, 
represent the disconnected contributions at this order. 
Here each vertex represent a factor of 
$\G^{12}$ which is defined as,
$$
\<\bar{O}_1(x)O_2(0)\>_{g_2^2}=-\G^{12}\lambda'\frac{\ln(x^2\Lambda^2)}
{(4\pi^2x^2)^{J+2}}.$$   
This quantity was first computed in \cite{CFHM} and given in
eqs. (\ref{G12}), (\ref{C12}) which show that each vertex scale with
a factor of $\frac{g_2}{\sqrt{J}}$. It is now clear that one can
reproduce
 all of the information contained in the scaling law of
(\ref{general}) 
by representing the correlators with these diagrams. For instance the
connected diagrams in Fig.2.a is $\O(g_2^2/J)$ hence vanishes in BMN
limit whereas the disconnected diagrams of the same order in Fig.2.b,c
and d scale as $g_2^2$ therefore they are finite 
because of the extra $J$ factor provided by the integration over 
the loop position. 

To make contact with light-cone
SFT we take this diagrammatic representation seriously with one
qualification: The matrix elements of the light-cone Hamiltonian should
correspond to
 the matrix $\G$ in the {\emph{string basis}}, not in BMN basis. 
As discussed in Appendix B, this matrix
element is obtained from $\G_{ij}$ with a unitary transformation, 
$\tilde{\G}\equiv U\G U^{\dg}$. Whatever the correct identification of
$U$ is, this transformation will not change the
scaling of $\G$ because it is independent of $J$. 
Therefore we can take the diagrams in Fig.2 seriously
as string theory diagrams\footnote{Of course one should replace strips
  in these 2D figures with tubes for closed SFT}, 
where the vertices $\G$ replaced with
$\tilde{\G}$ which scale in the same way as before. 
For instance, the vanishing of Fig.2.a implies that there is no 
double-double ``contact term'' that contributes to
$\<\psi_2|P^-|\psi_2\>$ at $\O(g_s^2)$. This
observation immediately generalize as, 
\begin{itemize}
\item {\emph{There are no $O(g_s^2)$, $i$-$i$ contact terms that contribute to $\<\psi_i|P^-|\psi_i\>$.}}
\end{itemize}  
However, these contact terms do give contributions at higher orders in
$g_s^2$. In other words, the suppression of the correlators in (\ref{general}) does
not imply the absence of physical information contained in these
quantities. They certainly yield non-zero contributions to
single-single loop corrections as illustrated in Fig.3. 

Let us also observe that the suppression of the
diagram in Fig.1.b implies that there is also no $2\to 1\to 2$ contribution
to this matrix element in string perturbation theory. This fact
generalizes as,

\begin{itemize}
\item {\emph{String theory processes where the number of internal
      propagations is less than $i$, do not contribute $\<\psi_i|P^-|\psi_i\>$.}}
\end{itemize}  

These assertions might seem strong, however one should note
the important assumptions that were made in the above
discussion. First of all the correspondence with GT, at the
perturbative level only holds when $\lm'\ll 1$ which translates into
the condition $\mu\gg 1$ in string theory. Therefore our discussion is
valid for large values of $\mu$. Secondly, the string 
amplitudes we consider involve a very particular class of external
string states, namely the states with only two excitations along $i=1,2,3,4$ directions 
(corresponding to scalar impurities in the BMN operators). Note however that our
discussion does not make any restriction to these particular two
scalar excitations in the {\emph{internal string states}}. 

\begin{figure}[htb]
\centerline{\epsfxsize=15cm\epsfysize=6cm\epsffile{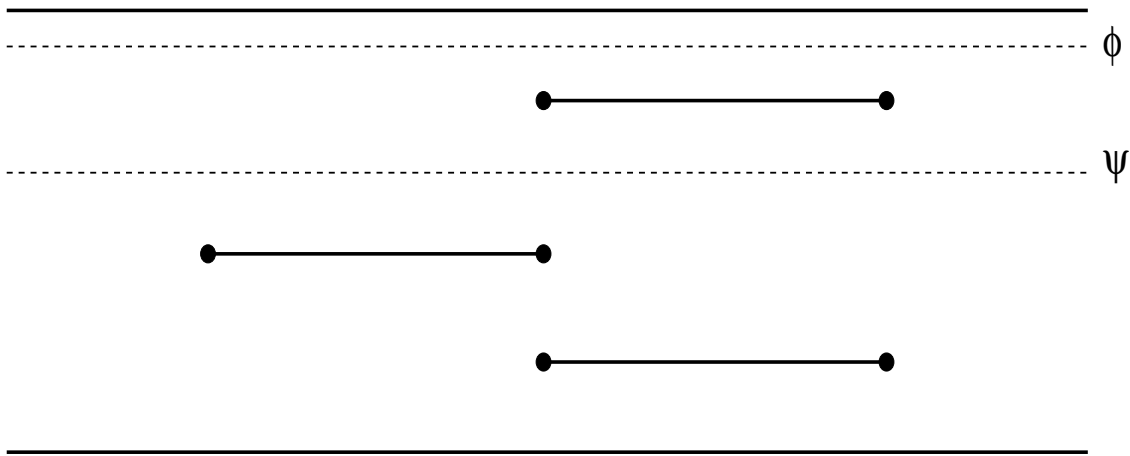}}
\caption{Connected contribution to double-triple correlator,
  $\<\bar{O}_{ny}O_{myz}\>$, does give non-zero contribution in 
  $1\to 1$ process. For example this diagram will show up in the computation
  of $\O(g^6)$ scale dimension of single-trace operators.}    
         \label{FIG3}
\end{figure}

\newpage

\section{Matrix elements of pp-wave Hamiltonian in 2-2 and 1-3 sectors}        
 
We will first compute the matrix elements of $P^-$ in the two-string sector by the method of \cite{Gomis}. 
Assuming the validity of the basis transformation $U_G$ that we
discussed in Appendix B, this will allow us to make 
a gauge theory prediction for SFT. Then, by the same method we will obtain the matrix elements in single-triple 
string sector. Let us briefly review the method. 

In Appendix B we presented a prescription to identify the string basis in field theory by transforming the 
basis of BMN observables with a real and symmetric transformation which renders the metric $G_{ij}$ diagonal. 
The conjecture is that, matrix elements of $P^-$ should be in correspondence with the matrix of $\O(\lm')$  
piece of the field theory correlators in the string basis. This is related to the same quantity in the 
old basis as, 
$$\tilde{\G}=U_{G}\G U_{G}^{\dg}.$$
We are interested in the $\O(g_2^2\lm')$ piece of $\tilde{\G}$. Using (\ref{U}), this reads \cite{Gomis},
\be\la{G2}
\tilde{\G}^{(2)}=\G^{(2)}-\half\{G^{(2)},\G^{(0)}\}-\half\{G^{(1)},\G^{(1)}\}+\frac{3}{8}\{(G^{(1)})^2,\G^{(0)}\}
+\frac{1}{4}G^{(1)}\G^{(0)}G^{(1)},
\ee
where the superscript denotes the order in $g_2$. Straightforward
algebra gives,
\bea\la{G22}
\tilde{\G}^{22}_{my;m'y'}&=&\G^{22}_{my;my'}-\half\left((\frac{m}{y})^2+(\frac{m'}{y'})^2\right)G^{22}_{my;m'y'}
-\half\left(G^{21}\G^{12}+\G^{21}G^{12}+G^{23}\G^{32}+\G^{23}G^{32}\right)_{my;m'y'}\nonumber\\
{}&+&\frac{3}{8}\left(G^{21}G^{12}+G^{23}G^{32}\right)_{my;m'y'}\left((\frac{m}{y})^2+(\frac{m'}{y'})^2\right)+
\frac{1}{4}\left(G^{12}G^{12}n^2+G^{23}\G^{33}G^{32}\right)_{my;m'y'}\nonumber
\eea
$\G^{22}$ and $G^{22}$ in the first two terms are $\O(g_2^2)$ pieces of the corresponding matrices 
and $\G^{33}$ in the last term is the $\O(1)$ piece. Repeated intermediate indices mean summing over 
all possible operators that may appear in that intermediate process \eg in the expression 
$G^{23}G^{32}$ one should sum over both BPS type {\emph{and}} BMN type triple trace operators. 
Remarkable simplifications occur, when one recalls that only non-vanishing contributions in the double-double 
sector comes from disconnected diagrams. A term like $G^{21}G^{12}$ and $G^{21}\G^{12}$ 
cannot be disconnected hence of $\O(1/J)$ and decouples in the BMN limit. Similarly one only keeps the 
disconnected contributions to $G^{22}$ and $\G^{22}$. All of the necessary ingredients to compute this expression 
except, 
\bd
\G^{33}=\left(\begin{array}{cc}
 \half(\frac{m}{y})^2\de_{m,m'}\de_{y,y'}(\de_{z,z'}+\de_{z,1-y-z'})& 0 \\
0 & 0 
\end{array}\right),
\ed
were presented in section 2. This matrix tells us that there is no
anomalous mixing among BPS type and between BMN
and BPS type triple-trace operators at the zeroth order in $g_2$. 

With the help of the decomposition identity (\ref{id1}), one 
obtains,    
\be\la{22matrix}
 \tilde{\G}^{22}_{my;m'y'}=g_2^2\de_{y,y'}\frac{y^2}{4}B_{m,m'}.
\ee
For the definition of matrix $B$, see Appendix C. 

A similar calculation yields the single-triple matrix element in the string basis as, 
\bea
\tilde{\G}^{13}_{m;m'y'z'}&=&\G^{13}_{m;my'z'}-\half\left(m^2G^{13}+G^{13}\G^{33}\right)_{m;m'y'z'} 
-\half\left(G^{12}\G^{23}+\G^{12}G^{23}\right)_{m;m'y'z'}\nonumber\\
{}&+&\frac{3}{8}\left(\G^{11}G^{12}G^{23}+G^{12}G^{23}\G^{33}\right)_{m;m'y'z'}
+\frac{1}{4}G^{12}_{m;ny''}(n/y'')^2G^{23}_{ny'';m'y'z'}.\nonumber
\eea
Again, repeated indices in the intermediate sums imply the inclusion of all possible operators of that given type. 
For instance in the term $G^{12}G^{23}\G^{33}$, one should use both BMN and BPS type double and triple operators in the 
intermediate process. A simplification occurs however when one notes that there is no $\og2$ contribution to $G^{23}$ 
and $\G^{23}$ for a BPS type double-trace operator, the lowest order non-zero contribution appearing at $\O(g_2^3)$. 
By repeated use of the decomposition identities (\ref{id2}),
(\ref{intdec1}) and (\ref{intdec2}) one obtains the amazingly simple expression, 
$$
\tilde{\G}^{13}_{m;m'y'z'}=\frac{g_2^2}{4}B^{13}_{m;m'y'z'}.
$$
Here $B^{13}$ is the contribution to $\G^{13}$ from non-nearest
neighbour interactions, given by (\ref{B13}). Thus we obtain the
following GT prediction for the matrix elements of $P^-$ in 1 string-3
string sector:
\be\la{13matrix}
\tilde{\G}^{13}_{m;m'y'z'}=\frac{g_2^2\lm'}{2\pi^3J}\frac{1}{(n-m/y)}\sqrt{\frac{z\tilde{z}}{y}}
\sin(\pi n z)\sin(\pi n \tilde{z})\sin(\pi n (1-y)).
\ee
Some comments are in order. First of all we note the striking
similarity of $\tilde{\G}^{22}$ and $\tilde{\G}^{13}$ to
$\tilde{\G}^{11}$ that was obtained in \cite{PSVVV}\cite{Gomis}:
$$\tilde{\G}^{11}_{m;m'}=\frac14 B_{m,m'}.$$
In the 2-2 sector this just follows from the disconnectedness of the
GT diagrams, hence the 2-2 matrix element just reduces to 1-1 case 
up to an overall factor $y^2$ therefore is hardly surprising. But our
result for the 1-3 matrix elements indicates the following generalization.   
As first computed by Vaman and Verlinde \cite{SBF} using SBF and then 
by Roiban, Spradlin and Volovich \cite{RSV} using rigorous SFT the 
matrix element $\tilde{\G}^{11}$ represents the ``contact term'' \ie 
the $\O(g_s^2)$ matrix element of $P^-$ between two single-string
states in the ST side. On the GT side, in all of the cases we
considered this matrix element is determined solely by the ``non-contractible''
contribution to $\G^{ij}$. It is tempting to conjecture that the
``non-contractible'' GT diagram encodes the information for the
$\O(g_s^2)$ contact term in PP-wave SFT. To check this conjecture one
should compute $\O(g_s^2)$ matrix element of $P^-$ between a single
and a triple-string state and compare with (\ref{13matrix}).   

\section{Discussion and outlook}

There are three main results in this manuscript. First of all, we
demonstrated that non-degenerate perturbation theory becomes invalid
at order $g_2^2$, as single-trace operators are degenerate with
triple-trace operators. This result casts some doubt on the previously
computed anomalous dimensions in \cite{CFHM}\cite{German2}. However we
conjectured that some particular eigenstates of the degenerate
subspace for finite $J$, tend to the modified BMN operators
$\tilde{O}_i$ in the BMN limit whose eigenvalues
coincide with the dimensions of $\tilde{O}_i$. Therefore the use of
non-degenerate perturbation theory can be justified for these
particular dilatation eigen-operators. This problem requires further
investigation and it will be interesting to explore new effects related
to this degeneracy problem in future.   

Our second main conclusion is the determination of the anomalous
dimensions of all multi-trace BMN operators that include two scalar
impurity fields in terms of the single-trace anomalous dimension. We
proved this interesting result to order $g_2^2\lm'$ but the fact that 
connected field theory diagrams are suppressed also at higher orders 
in $g_2$ suggests that the conclusion holds 
at an arbitrary level in perturbation theory. (Of course, one has to
first establish the validity perturbation theory at higher orders.) These
predictions for the eigenvalues of $P^-$ are basis independent and
therefore provide a firm prediction for SFT. 
It would be interesting to understand the string theory mechanism that is analogous
to the BMN suppression that leads to vanishing of connected field
theory diagrams. A natural next step in this analysis is to
consider the anomalous dimensions of BMN operators that include higher
number of impurities. We believe that suppression of the disconnected
GT diagrams will lead to remarkable simplifications also in that problem.  

Finally, we obtained predictions for the matrix elements of the light-cone
Hamiltonian in 2-2 and 1-3 string sectors. We emphasize that these
predictions are sensitive to the way the string basis in GT is
identified, unlike the predictions of section 3 for the
{\emph{eigenvalues}} of $P^-$. We fixed the basis with the assumption
that the form of the basis transformation at $\og2$ is also valid at
$\O(g_2^2)$. Although this assumption passed a non-trivial test in
predicting the correct $\O(g_2^2)$ contact term of single-string
states \cite{RSV}, there is no obvious reason to believe its validity
for instance in the single-triple sector. Thus, our predictions can
also be used as a test of the basis identifications of either \cite{PSVVV} or 
\cite{Gomis} which are equivalent to each other at $\O(g_2^2)$.

\centerline{\bf Acknowledgments}

It is a pleasure to thank Neil Constable, Dan Freedman, Matt Headrick,
Charlotte Kristjansen, Mark Spradlin and Anastasia Volovich for useful discussions. 
This work is supported by funds provided by the D.O.E. under 
cooperative research agreement $\#$DF-FC02-94ER40818. 

\newpage

\appendix    

\section{Disconnectedness of GT correlators}

In this appendix we will derive eq. (\ref{general}). A study of the
corresponding Feynman diagrams suffice to obtain the
leading order scaling of a generic correlator with $g_2$ and $J$. 
Dependence on $g_2$ of a correlator is fixed by power of $N$ in a Feynman diagram. This can be determined 
either by direct evaluation of the traces over the color structure (all fields are in adjoint rep. 
in ${\cal N}=4$ SYM) or by loop-counting. Since we are interested in the leading order $g_2$ dependence, 
the latter is easier. Explicit $J$ dependence is determined by working out 
the symmetry factors in a Feynman diagram. As a warm-up consider the {\emph{free}} extremal correlation function, 
$$C^{i,1}=\<\bar{O}_n^{J}:O_m^{J_{1}}O^{J_{2}}\cdots O^{J_{i}}:\>.$$ Leading order diagram drawn on a plane is shown in Fig.4. 
Taking the normalization factor $\sim 1/N^{J+2}$ into account, trivial loop counting teaches us that, 

\begin{figure}[htb]
\centerline{ \epsfysize=8cm\epsffile{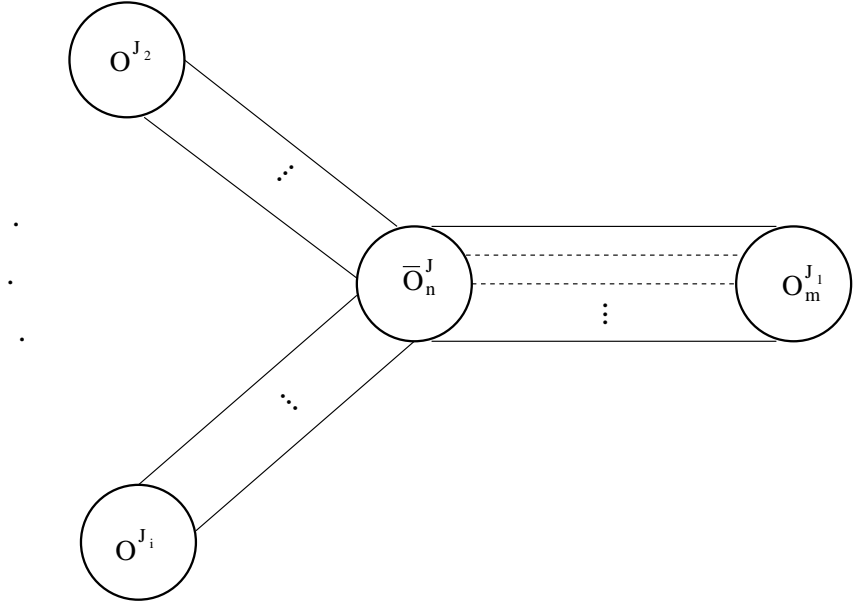}}
\caption{A typical planar contribution to $C^{1,i}$. Circles represent
  single-trace operators. Dashed lines
  denote impurity fields. $Z$ lines are not shown explicitly and
  represented by ``$\cdots$''.} 
         \label{FIGA1}
\end{figure}

\be\la{armageddon} 
C^{i,1}\propto\frac{1}{N^{i-1}}=\frac{g_2^{i-1}}{J^{2i-2}}.
\ee
Now, consider the combinatorics in Fig.4 to determine the power of $J$. Planarity requires Wick contraction of $O^{J_i}$'s into 
$\bar{O}_n^{J}$ as a whole. Fix the position of, say $O^{J_2}$ in $\bar{O}_n^{J}$. Then one has to sum over positions of 
other $O^{J_i}$ operators for $i>2$ within $\bar{O}_n^{J}$ obtaining a factor of $J^{i-2}$. 
There is a phase summation over positions of $\phi$ and $\psi$ impurities in $\bar{O}_n^{J}$, 
giving a factor of $J^2$. Cyclicity of $O^{J_i}$, $i>1$, provides a factor of $J^{i-1}$. Taking into account 
the $J^{-(i+1)/2}$ suppression from the normalization and $\O(J^{-2i+2})$ suppression in (\ref{armageddon}), 
we learn that,
\be\la{extremal}
C^{i,1}\sim\frac{g_2^{i-1}}{J^{(i-1)/2}}.
\ee

Next task is to obtain similar information for a general, non-extremal {\emph{free}} correlator in (\ref{corr}). 
Without loss of generality, one can assume $j\le i$. There are various connected and disconnected diagrams with 
different topology. Since the results for disconnected contributions will recursively be included in 
the fully connected pieces for smaller $i$ and $j$, it suffices to consider the fully-connected contribution to 
(\ref{corr}). We first ask for the dependence on $g_2$ for the leading order (planar) fully connected diagram. As an example, 
a list of all distinct topologies for fully connected $i=3$, $j=2$ correlator is shown in Fig.5. It is immediate to see that 
conservation of number of legs for each operator in the correlator
(for each node in Fig.5) requires that {\emph{all planar 
fully-connected diagrams have same $g_2$ power}} irrespective of the
topology (here, by topology we refer to different type of diagrams that
are exemplified in Fig.5, not the order in $g_2$). Then, it is sufficient to count the loops in a 
connected diagram that is the simplest for loop counting
purposes. This simplest diagram is shown in Fig.6. Each outer leg in Fig.6 represent a bunch of 
$J_s$ propagators (inner line has $J_1-J_{i+2}-\cdots-J_{i+j}+2$
propagators). Drawn on a plane, this means that there are a total of 
$(J+2)-(i+j-1)+1$ loops in Fig.6, including the circumference loop. Finally, a factor of $N^{J+2}$ from normalizations and  
we obtain that, 
\be\la{emperyalistpatates}
C^{i,j}\propto\frac{1}{N^{i+j-2}}=\frac{g_2^{i+j-2}}{J^{2(i+j-2)}}.
\ee

Apart from the dependence on $J$ coming from singling out the $g_2$
dependence as above, there are additional contributions from the
combinatorics and normalizations. Determination of the power of $J$ from the combinatorics works much as 
in the case of $C^{1,i}$. Fixing position of one $O^{J_i}$ inside another operator that it connects to, we are left with sum 
over position of $i+j-3$ operators. This reasoning holds only for
tree-type diagrams like in the first diagrams shown in Fig.5.a and
Fig.5.b.  

\begin{figure}[htb]
\centerline{ \epsfysize=6cm\epsffile{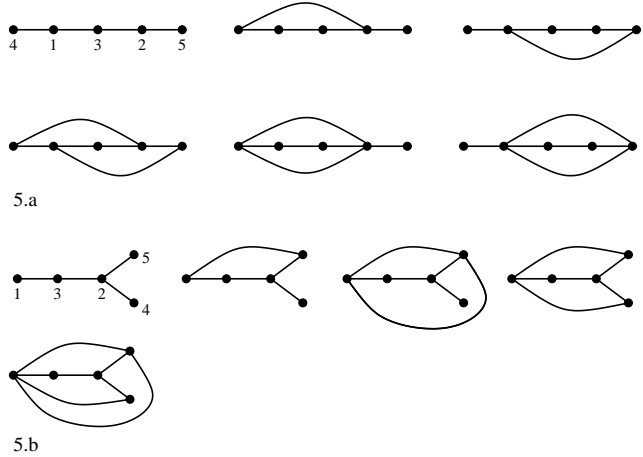}}
\caption{All distinct topologies of planar Feynman diagrams that
  contribute to $\<:\bar{O}^{J_1}_n\bar{O}^{J_2}::O^{J_3}_mO^{J_4}O^{J_5}:\>$. 
Nodes represent the operators while solid lines represent a bunch of
  $Z$ propagators. Line between the nodes 1 and 3 also include two scalar
  impurities $\phi$ and $\psi$. 
  All other topologies are obtained from these two classes by permutations among
  3,4,5 and 1,2 separately. Other planar graphs are obtained from these by moving the
  nodes within the solid lines without disconnecting the diagram. For
  example 4 in the first diagram can be moved within the solid line
  1-3.}       
\end{figure}

But it is not hard to see that the combinatorial factor for diagrams
involving loops \eg second diagram in Fig.5.
is also equal to $i+j-3$. This is because for each factor 
that one loses from the sum over positions of $O_{J_i}$ because of
the appearance of a loop, one gains a compensating 
factor of $J$ for the loop summation. Also, cyclicity of BPS type operators
within (\ref{corr}) provides a factor of $J^{i+j-2}$. Finally, 
including the powers of $J$ coming from the normalizations and (\ref{emperyalistpatates})
 we arrive at the general result,    
\be\la{gen}
C^{i,j}\sim\frac{g_2^{i+j-2}}{J^{(i+j)/2-1}}.
\ee

Note that this is for the connected contribution to
(\ref{corr}). To obtain the $g_2$ and $J$ dependence of disconnected
contributions, one simply uses (\ref{gen}) for smaller $i$ and/or $j$
which shows that disconnected diagrams have lower powers in $g_2$
and they are less suppressed by a power of $J$. For example a
disconnected contribution to $C^{ij}$ where the process $i\to j$ is
separated into two disconnected processes, the scaling would be,
$$C^{i,j}\sim\frac{g_2^{i+j-2}}{J^{(i+j)/2-2}}.$$ 

\begin{figure}[htb]
\centerline{ \epsfysize=6cm\epsffile{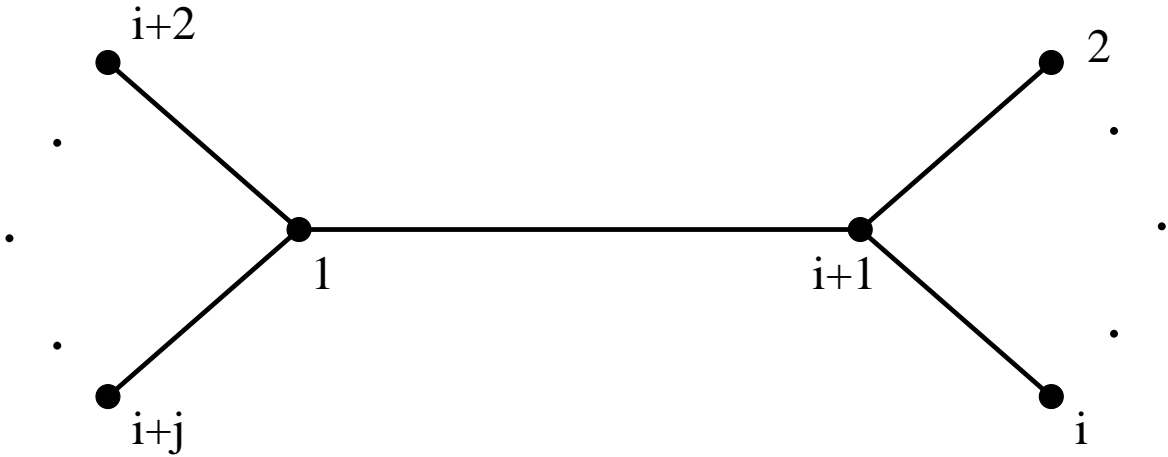}}
\caption{Simplest connected tree diagram for the loop counting
  purposes. The solid line between 1 and $i+1$ includes $\phi$ and
  $\psi$.}
\end{figure}

With a little more effort one can show that, 
\begin{theo}\la{yelkovan}
For scalar impurity BMN operators $\O(g_{YM}^2)$ 
interactions will not change the scaling law of (\ref{gen}) at all.
\end{theo}
 Let us outline the proof shortly. As a first step, one can show that
 the only interactions
involved in {\emph{scalar impurity}} BMN operators are coming from
F-terms in ${\cal N}=4$ SYM lagrangian. F and D type interaction terms
written in ${\cal N}=1$ component notation reads, 
\be\la{FD}F \propto (f^{abc}\bar{Z}^i_b Z^i_c)^2,\;\;\;
D \propto f^{abc}f^{ade}\epsilon_{ijk}\epsilon_{ilm}Z^j_bZ^k_c\bar{Z}^l_d\bar{Z}^m_e.\ee
Here, $a,\dots$ are the color indices while $i,\cdots$ denote
flavor. Note that when one specifies the orientation in a scalar
propagator $Z^i\bar{Z}^i$ as from $Z$ to $\bar{Z}$, these quartic
vertices can be represented as in Fig 8. In \cite{Skiba} it was shown
that, correlation functions of BPS type multi-trace operators, 
\be\la{multi}\Tr(Z^J_1)\cdots\Tr(Z^J_r)\ee
do not receive any radiative corrections. To see this one first notes
that F-type quartic vertex vanishes when fields are all have the same
flavor. Secondly one discovers that contribution of D-type quartic
vertex exactly cancels out the contributions from self energies and
gluon exchange \cite{Skiba}. 

\begin{figure}[htb]
\centerline{ \epsfysize=8cm\epsffile{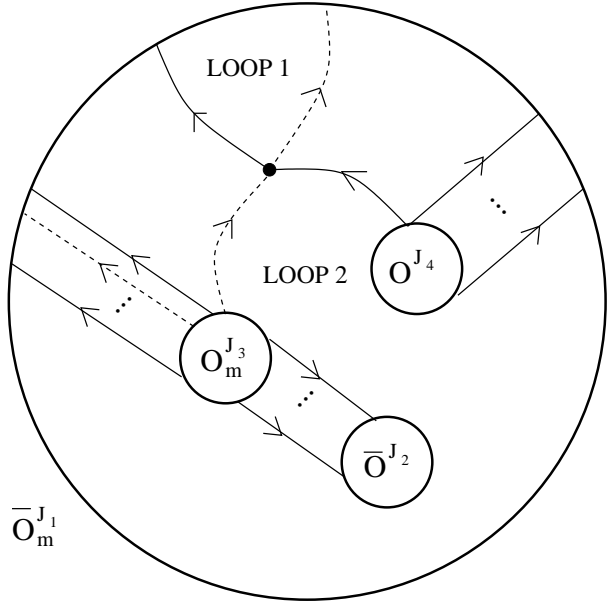}}
\caption{A quartic interaction in a typical diagram introduce two
  interaction loops. Here, loop 1 is contractible while loop 2 is
  non-contractible, therefore this diagram represents a
  semi-contractible interaction.}
\end{figure}

Now consider replacing some of the BPS
operators in (\ref{multi}) with BMN operators, (\ref{BMNop}). Since the scalar
impurities in BMN operators are distinguished from $Z$ fields by their
flavor, F-terms are now allowed. However, unlike F-type interactions 
D-term quartic vertex, gluon exchange and self energies are all flavor
blind, therefore one can replace the $\phi$ and $\psi$ impurities with 
$Z$ fields for the sake of studying possible contributions from
these interactions. After this replacement the phase sum over the
position of impurities in $O_n^J$ becomes trivial and factors out of the
operator, hence BMN operators reduce to BPS type operators times an
overall phase factor. Therefore the theorem of \cite{Skiba} for 
BMN type multi-trace operators becomes, 

\begin{theo}\la{skiba}
The only radiative corrections to n-point functions of multi-trace BMN
operators come from F-type interactions.
\end{theo}
  
Second step in the proof of theorem 1 is the classification of 
topologies of Feynman diagrams with one F-term interaction. 
Any $\O(\lm')$ interaction that one inserts in
(\ref{corr}) introduces two ``interaction loops'' on the plane
diagrams. A generic example is shown in Fig.7. According to the
contractibility of these interaction loops one can classify planar
F-term interactions as, 
\begin{enumerate}   
\item {\emph{Contractible}}: Both interaction loops are contractible,
\item {\emph{Semi-Contractible}}: Only one of the loops is contractible,
\item {\emph{Non-Contractible}}: None of the loops are contractible.
\end{enumerate}
Requirement of contractibility means that two incoming lines and two
outgoing lines in the F-term vertex of Fig.8 i) belong to the same operator and ii) 
adjacent to each other when drawn on a plane. 
Now, we note that this classification of interactions would hardly make any difference if
we were not dealing with BMN type operators which involve a
non-trivial phase summation over the position of the
impurity. Structure of the F-term interactions in (\ref{FD}) makes it
clear that interactions of adjacent lines yield a phase factor
$$1-e^{\frac{2i\pi n}{J_1}} \sim \frac{1}{J}.$$ 
Therefore we learn that the phases in BMN operators provide a factor
of $1/J^2$ for ``contractible'' interactions, $1/J$ for
``semi-contractible'' interactions, $1/J^0$ for ``non-contractible''
interactions. 

\begin{figure}[htb]
\centerline{ \epsfysize=4cm\epsffile{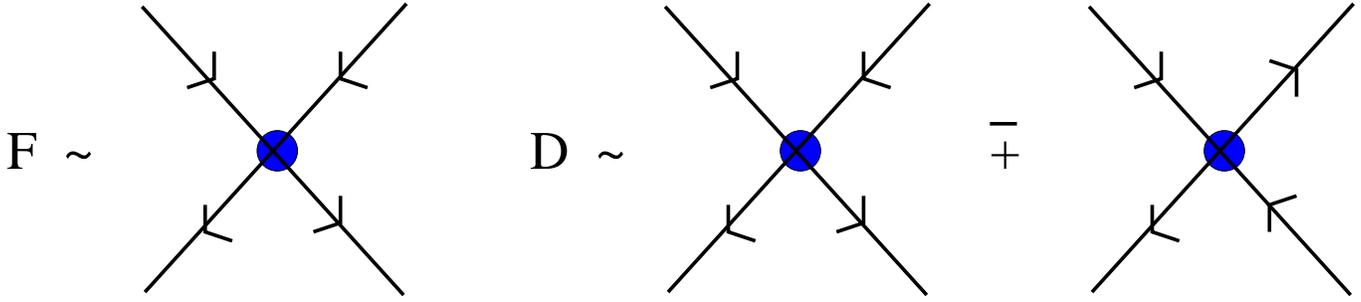}}
\caption{Orientations of F-term and D-term quartic vertices.}
\end{figure}

As the last step in our proof of theorem 1, let us show that non-contractibility of 
each interaction loop supplies another factor of $1/J$. It should 
be clear from above requirements for contractibility that there are 
two distinct situations that non-contractibility of an interaction loop
can arise:
\begin{enumerate}   
\item the incoming (or outgoing) lines in Fig.8 belong to different 
operators within a multi-trace operator {\emph{or}} 
\item belong to the same operator but are not 
adjacent to each other.
\end{enumerate}
Let us now recall that among various contributions to the power of $J$
in free correlators, there is a combinatorial factor of $J^{i+j-3}$
coming from summing over positions of the insertions of $O^{J_i}$
operators inside $\bar{O}^{J_i}$'s. In case 1 above, clearly, one of these
position sums will be missing, hence a suppression by $1/J$. 
On the other hand case 2 can only arise
in a situation where there is at least one operator inserted in
between the incoming or (outgoing) lines which take place in
the interaction. Since the position of this inserted operator is
required to have a fixed position in between the interacting legs one
also arrives at a suppression by $1/J$. These two situations are 
illustrated in an example of $G^{3,2}$ in Fig.9. 

\begin{figure}[htb]
\centerline{\epsfxsize=15cm\epsfysize=6cm\epsffile{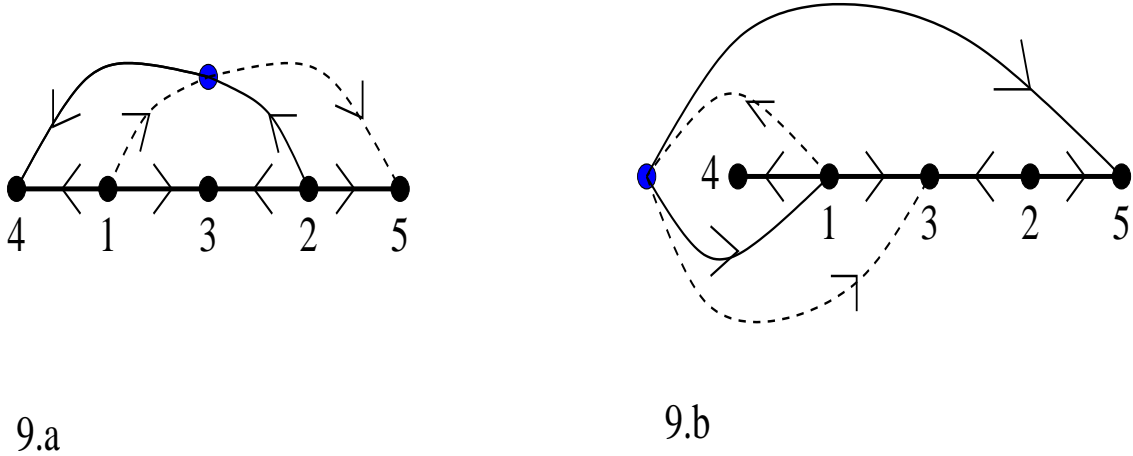}}
\caption{In {\bf a} both of the interactions loops are non-contractible
  due to case 1: incoming and outgoing line pairs of the quartic
  vertex connect to different operators. In {\bf b} one interaction
  loop is non-contractible due to case 1 the other due to case 2.} 
\end{figure}

When combined with the powers of $J$ coming from the phase factors that we
described above, we see that they compensate each other and one gets a
universal factor $1/J^2$ for all of the different topologies in an
F-term interaction, namely contractible, semi-contractible and
non-contractible. Finally note that all interactions come with a
factor of $g_{YM}^2N$. Combined with $1/J^2$ this yields $\lm'$ and
therefore we concluded the proof of theorem 1: One-loop radiative
corrections to (\ref{corr}) is of the form,
$$C^{i,j}\sim\frac{g_2^{i+j-2}}{J^{(i+j)/2-1}}\lm'.$$

The reason that this theorem might fail in case of BMN operators with
non-scalar impurities is that D-term interactions might give non-vanishing
contributions (see \cite{vector}). 

\section{Field theory basis change at finite $g_2$}

Apart from the purely field theoretical task of determining the eigenvalues and eigen-operators of 
$\Delta-J$, one can ask for the identification of the string basis in field theory for non-zero $g_2$ 
\cite{PSVVV}\cite{Gomis}. For non-zero $g_2$, the inner product in field 
theory becomes non-diagonal in the original basis of BMN where there is an explicit identification 
of $n$-string states with $n$-trace operators. On the other hand, to identify the ``string theory basis'' in 
field theory one should require the field theory inner product to be diagonal for all $g_2$. 
However, this requirement alone {\emph{does not}} uniquely specify the necessary basis change from 
BMN basis to string theory basis for finite $g_2$. One always has the freedom of 
performing an arbitrary unitary transformation.    

In the recent literature, two independent but compatible approaches for the identification of string basis 
were presented. In the string bit formalism (SBF), it is possible to 
capture the kinematics and dynamics of gauge theory amplitudes by the discretized theory of bit strings.
\footnote{Strictly speaking this has been shown only at $\O(g_2^2\lm')$ and only 
for scalar impurities\cite{SBF}.} According to the conjecture of \cite{PSVVV}, 
in the string bit language, the basis 
transformation which takes from BMN basis to string basis for all $g_2$ reads, 
\be\la{sbt}
|\tilde{\psi_i}\>=\left(e^{-\frac{g_2}{2}\Sigma}\right)_{ij}|\psi_j\>\equiv 
\left(U_{\Si}\right)_{ij}|\psi_j\>\,
 \ee
where $\psi_n$ denotes an $n$-string state. Here $\Si$ is the sum over all distinct 
transpositions of two string bits, 
$$\Si=\frac{1}{J^2}\sum_{mn}\Si_{(mn)}.$$ $\Si$ has the effect of a single string splitting or joining
, \ie it can map an $i$ string state into an $i\pm 1$-string state. Note that the transformation 
matrix  $e^{-\frac{g_2}{2}\Sigma}$ is real and symmetric. 

Another method \cite{Gomis} which leads an identification of
 the string basis is simply to find the transformation $U$ which
 diagonalizes  the matrix of inner products between BMN operators, order by order in $g_2$. 
In the {\emph{free}} theory, we define the following matrix, 
$$G_{ij}=\<\bar{O}_jO_j\>.$$
Here $n$ is a collective index for a generic $n$-trace BMN operator. One identifies the 
basis transformation, $U$ by requiring that $G$ is diagonal in the new, ``string basis'': 
$$U_{ik} G_{kl}U_{lj}^{\dg}=\de_{ij},\;\;\;\tilde{O}_i\equiv U_{ij}O_j.$$
As in the SBF basis change $U$ is specified up to an arbitrary unitary transformation. 
One can fix this freedom by requiring that $U$ is real and symmetric. Call this transformation, $U_G$. 
Then the solution of the above equation up to $\O(g_2^2)$ reads, 
\be\la{U}
U_G={\mathbf{1}}+g_2 (\half G^{(1)}) + g_2^2(-\half G^{(2)}+\frac{3}{8}(G^{(1)})^2)+\cdots
\ee    
where $G^{(i)}$ denotes $\O(g_2^i)$ piece of the metric. 

The requirement of reality and symmetry completely fixes the freedom in the choice of 
the transformation. It was independently shown in \cite{PSVVV} and \cite{Gomis} that this 
simple choice leads to an agreement with string theory calculations. In particular the inner 
product of a single and double trace operator in the {\emph{interacting}} theory in this basis, 
agrees with the cubic string vertex. Recently, \cite{RSV} also gives evidence for an 
 agreement between the $\O(g_2^2\lm')$ eigenvalue of $\Delta-J$ and the matrix element of light cone 
Hamiltonian in single string sector.
Despite the agreement at this order, there is no reason to believe that this simple choice 
should hold at higher orders in $g_2$ {\emph{or}} for higher trace multi-trace operators. 

Here, we will confine ourselves to checking the compatibility of this basis transformation 
with free field theory correlators at $g_2^2$, including the effects of triple-trace 
operators. As a bonus we obtain new deconstruction identities decomposing some 
multi-trace inner products in terms of smaller multi-trace inner products. 
Since the requirement of reality and symmetry of the basis transformation matrix
completely fixes the definition of the new basis, the two prescriptions described above 
should be equivalent. Equating various matrix elements of $U_G$ and $U_{\Si}$ we will 
obtain identities involving free field theory correlators which are subject to 
should be check in field theory. 

The first non-trivial requirement is coming from the single-trace operators at $\O(g_2^2)$, 
$$\<\psi_1| \tilde{\psi}_1\>_G = \<\psi_1| \tilde{\psi}_1\>_{\Si}.$$    
LHS is, (we suppress indices labeling a multi trace operator), 
$$\<\psi_1|U_G|\psi_1\>= 1+g_2^2(-\half G^{11}+\frac{3}{8}G^{12}G^{21}).$$ $G^{11}$ is the 
torus level, free single-single correlator with the space-time dependence removed. Similarly 
$G^{12}$ is $\og2$ level single-double correlator which is presented
in eq. (\ref{C12}). RHS reads,
$$\<\psi_1|1+\half(-\frac{g_2^2}{2})^2\Si^2|\psi_1\> = \frac{g_2^2}{8} G^{12}G^{21},$$
where we used the fact that $\Si$ changes the string number by $\pm 1$. Hence we obtain, 
\be\la{id0}
G^{11}_{nm}=\half\sum_{i}G^{12}_{n;i}G^{12}_{i;m}.
\ee
 Here $i$ is a collective index 
labeling either of the two types of double-trace operators which can appear in the intermediate 
process \ie $i=\{ny\}$ for BMN double-trace and $i=y$ for BPS double-trace. 

This identity indeed holds in the gauge theory as first shown in $\cite{Huang}$. 
With the inverse reasoning we can view this simple calculation as the {\emph{derivation}} of 
this non-trivial sum formula. As a next application let us derive identities involving 
triple trace operators. We require, 
$$\<\psi_3|\tilde{\psi}_1\>_{G}= \<\psi_3|\tilde{\psi}_1\>_{\Si}.$$ 
Much as above, $O(g_2^2)$ term in LHS is, 
$$-\half G^{13} +\frac{3}{8}G^{12}G^{23}$$ and the RHS is, 
$$\frac{g_2^2}{8}\<\psi_3|\Si^2|\psi_1\>=\frac{g_2^2}{8}G^{12}G^{23}.$$ We learn that
$$G^{13}=\half G^{12}G^{23}.$$
Notice that ``2'' in the intermediate step can not be a BPS double-trace operator, because 
the lowest order $G^{23}$ between a BPS double-trace and a BMN triple trace is at $\O(g_2^3)$.
\footnote{It is easy to see (either by using trace algebra or by counting the loops in Feynman diagrams) 
that there exists only {\emph{disconnected}} $\O(g_2)$ contributions 
to any double-triple correlator where the 2-3 
correlator separates as 1-2, and 1-1. This obviously cannot 
happen for a correlation function of a BPS double-trace operator and a BMN triple-trace operator.} 
Therefore we arrive at the formula, 
\be\la{id1}
G^{13}_{m;nyz}=\half\sum_{py'}G^{12}_{m;py'}G^{23}_{py';nyz}.
\ee
The single-triple correlation function $G^{13}$ is computed in
Appendix D. 
Again this identity is subject to check by a direct field theory computation. This is done 
in Appendix C and (\ref{id1}) passes the test. A similar calculation with the requirement 
$\<\psi_2|\tilde{\psi}_2\>_G=\<\psi_2|\tilde{\psi}_2\>_{\Si}$, one reaches at another useful 
identity, 
\be\la{id2}
G^{22}_{my;m'y'}=\half\left(\sum_n G^{12}_{my;n}G^{12}_{n;m'y'}+\sum_i G^{23}_{my;i}G^{23}_{i;m'y'}\right).
\ee
Here $i=\{ny''z''\}$ or $i=\{y''z''\}$ for BMN and BPS triple-trace operators. An expression for the free $\O(g_2^2)$ 
 double-double correlator, $G^{22}$, is given in eq. (\ref{Cii2})  
We also checked this by direct computation in Appendix C. We emphasize that these are highly non-trivial 
identities viewed as representation of a trigonometric function, say
 $G^{13}$ in (\ref{G13}) as 
an infinite series of products of simpler functions. In comparison to (\ref{id0}) the non-triviality 
comes from the fact that summands are trigonometric functions rather that rational functions as in the RHS
of (\ref{id0}). These identities will prove extremely useful for the computations of 
the next sections. Finally we note immediate generalizations, 
\bea
G^{m,m+2}&=&\half G^{m,m+1}G^{m+1,m+2}\nonumber\\
G^{mm}&=&\half\left(G^{m,m-1}G^{m-1,m}+G^{m,m+1}G^{m+1,m}\right).\nonumber
\eea

\section{Sum formulas}

Let us first reproduce the matrices $A_{mm'}$ and $B_{mm'}$ that
appear in the $\O(g_2^2)$ and  $\O(g_2^2\lm')$ pieces of the
single-trace two-point function. These were defined in
\cite{Constable}:
\be\la{AA}
A_{m,n}=\left\{\begin{array}{ll}
\frac1{24}, &  m=n=0; \\
0, & m=0, n\ne 0\,\, {\mathrm{ or }},\,\ n=0, m\ne 0; \\
\frac1{60} - \frac1{6u^2} + \frac7{u^4}, & m=n\ne 0; \\
\frac1{4u^2}\left(\frac13+\frac{35}{2u^2}\right), & m=-n\ne 0; \\
\frac1{(u-v)^2}
\left(\frac13+\frac4{v^2}+\frac4{u^2}-\frac6{uv}-\frac2{(u-v)^2}\right), 
& {\mathrm{all\,\, other\,\, cases}}
\end{array}\right.
\ee
and
\be\la{BB}
4\pi^2 B_{m,n}=\left\{\begin{array}{ll}
0, &  m=n=0; \\
\frac13+\frac{10}{u^2}, & m=n\ne 0; \\
-\frac{15}{2u^2}, & m=-n\ne 0; \\
\frac6{uv}+\frac2{(u-v)^2}, & {\mathrm{all\,\, other\,\, cases}}
\end{array}\right.
\ee
where $u=2\pi m, u=2\pi n$.

The rest of this appendix outlines the computation of non-trivial summations that
appear along the computations in sections 2, 4, 6 and the previous appendix. We will first
prove that (\ref{id1}) and (\ref{id2}) indeed hold in GT. These were
obtained in the previous appendix simply by
equating the basis transformations $U_{\Si}$ and $U_G$. A second task
is to derive (\ref{sum11}) of section 4. Finally we will define and
evaluate the last term in (\ref{13matrix}) of section 5. 

We reproduce (\ref{id1}) here for completeness: 
\be\la{id11}
G^{13}_{m;nyz}=\half\sum_{py'}G^{12}_{m;py'}G^{23}_{py';nyz}.
\ee
LHS of (\ref{id11}) is computed in Appendix D and presented in
eq. (\ref{C13}). $G^{23}$ that appear on the RHS gets $\og2$
contributions only from disconnected diagrams as $2\to 3$ decouples
as, $1\to 2$ and $1\to 1$. This quantity is also computed in Appendix
D, result is given in (\ref{C23}). One gets two contributions to RHS
of (\ref{id11}) from first and second pieces in (\ref{C23}). Second
contribution gives, 
\be\la{r1}
\half\int_0^1\ud y'\sum_{p=-\infty}^{\infty}G^{12}_{m,py'}\frac{g_2}{\sqrt{J}}\de_{np}\de_{yy'}\sqrt{(1-y)z\tilde{z}}
=\half\frac{g_2}{\sqrt{J}}\sqrt{(1-y)z\tilde{z}}G^{12}_{m,ny}.
\ee
First piece in $G^{23}$ gives, 
\be\la{r2}
\half\sqrt{\frac{z\tilde{z}}{y}}\int_0^1\ud y'
y'^3(\de_{y',y+z}+\de_{y',y+\tilde{z}})\frac{\sin^2(m\pi
  y')}{\pi^4}\left\{\sum_{p=-\infty}^{\infty}\frac{sin^2(p\pi
    y/y')}{(p-my')^2(p-ny'/y)^2}\right\}.
\ee
We will now describe the evaluation of the sum in this expression. Let
us separate the sum into two pieces as,
\be\la{sumtot}
S=S_1+S_2\equiv \sum_{p=-\infty}^{\infty}\frac{1}{(p-a)^2(p-b)^2}
-\sum_{p=-\infty}^{\infty}\frac{\cos^2(p x)}{(p-a)^2(p-b)^2},
\ee
where we defined,
$$ x\equiv\pi y/y',\;a\equiv my',\;b\equiv ny'/y.$$ 
$S_1$ is easy to evaluate (can be done with a computer code) and the
result is,
\be\la{sum1}
S_1=\frac{1}{(a-b)^3}\left(2\pi\cot(\pi a)-2\pi\cot(\pi
  b)+\frac{(a-b)\pi^2}{\sin^2(\pi a)}+\frac{(a-b)\pi^2}{\sin^2(\pi
    b)}\right).
\ee
It is not possible to evaluate $S_2$ neither with a well-known computer
program nor it can be found in standard tables of infinite series
(like Gradhsteyn-Rhyzik or Prudnikov). To tackle with it we reduce it
into a product of two sums as,
\bea
S_2&=&\frac{d^2}{dadb}\left(\sum_{p=-\infty}^{\infty}\frac{\cos(px)}{(p-a)}\right)
\left(\sum_{r=-\infty}^{\infty}\frac{\cos(rx)}{(r-b)}\de_{pr}\right)\nonumber\\
{}&=& \frac{d^2}{dadb}\frac{1}{2\pi}\int_0^{2\pi}\ud t\left(\sum_{p=-\infty}^{\infty}\frac{\cos(px)e^{ipt}}{(p-a)}\right)
\left(\sum_{r=-\infty}^{\infty}\frac{\cos(rx)e^{-irt}}{(r-b)}\right),\nonumber
\eea
where in the second step we used the integral representation of
$\de_{pr}$. Expanding the exponentials in terms of $\cos$ and $\sin$
we now reduced the sum into sums of the following form,
\be\la{daphne}
\sum_{p=-\infty}^{\infty}\frac{\cos(p(x\pm
  t))}{p-a}=-\frac{1}{a}+2a\sum_{p=1}^{\infty}\frac{\cos(p(x\pm t))}{p^2-a^2}=
-\frac{1}{a}+2a f_m(x\pm t,a).
\ee
We can read off the function $f_m(z,a)$ from \eg\cite{GR},
\be\la{f}
f_{m}(z,a)=\frac{1}{2a^2}-\frac{\pi}{2a}\frac{\cos(a(z-(2m+1)\pi))}{\sin(\pi
  a)},
\ee  
where $m$ is an integer defined as, $2\pi m\le z\le 2\pi(m+1)$. With
the given information it is straightforward to evaluate these 
sums. Integrating over $t$ and combining with $S_1$ in (\ref{sum1}),
one gets,
\bea
\sin^2(\pi a) S&=& \frac{\pi}{2(a-b)^3}\left[\sin(2\pi
  a)-\sin(2ax)-\sin(2a(\pi-x))\right]\nonumber\\
{}&+&\frac{\pi}{(a-b)^2}\left[x\sin^2(\pi
  a)+x\sin^2(a(\pi-x))+(\pi-x)\sin^2(ax)\right]\nonumber.
\eea
We insert this expression into (\ref{r2}), carry out the trivial
integration over $y'$. Using 
the definitions of $x$, $a$ and $b$ given above one
gets,
\bea
&&-\frac{\pi}{2(m-k)^3}\left\{\sin(2\pi mz)+\sin(2\pi
  m\tilde{z})+\sin(2\pi my)\right\}+\frac{\pi^2}{(m-k)^2}y
(\sin^2(\pi mz)+\sin^2(\pi m\tilde{z}))\nonumber\\
&&+\frac{\pi^2}{2(m-k)^2}(1-y)\sin^2(\pi my),\nonumber
\eea
where $k=n/y$. 
Comparison of this expression with (\ref{C13}) shows that this
expression equals,
$$G^{13}_{m;nyz}-\frac{g_2}{2}\sqrt{\frac{z\tilde{z}(1-y)}{J}}G^{12}_{m,ny}$$
Adding up to this the first contribution in (\ref{r1}) we proved 
(\ref{id11}). 

Now, let us move on the proof of the second decomposition identity,
(\ref{id2}) that we reproduce here,
\be\la{id22}
G^{22}_{m,y;m',y'}=\half\left(\sum_n G^{12}_{m,y;n}G^{12}_{n;m'y'}
+\sum_i G^{23}_{m,y;i}G^{23}_{i;m'y'}\right).
\ee
As mentioned before, there are disconnected
and connected contributions to both LHS and RHS of this
equation. Since connected contributions differ from the disconnected ones
by a factor of $1/J$, one should match $\O(1)$ and $\O(1/J)$ pieces on
both sides separately. Here we will present the equality of $\O(1)$
parts of LHS and RHS and leave the question of $\O(1/J)$ pieces for
future. We did not need $\O(1/J)$ terms anywhere in our computations. 
$\O(g_2^2)$ disconnected contribution to $G^{22}$ is just 
$$\<\bar{O}^{J_1}_nO^{J_3}_m\>_{g_2^2}\<\bar{O}^{J_2}O^{J_4}\>_{g_2^0}$$
plus
$$\<\bar{O}^{J_1}_nO^{J_3}_m\>_{g_2^0}\<\bar{O}^{J_2}O^{J_4}\>_{g_2^2}.$$
All required terms here were already computed in the literature (see
\cite{German1}\cite{Constable}) and the total result is, 
$$g_2^2\left(y^4A_{mm'}+\frac{\de_{mm'}}{24}(1-y)^4\right).$$

Turning to the RHS of (\ref{id22}) now, we first note that $2\to 1\to 2$
process can not be disconnected hence does not contribute at this
order. Evaluation of the second term in RHS is straightforward by
using (\ref{C23}) that is derived in the next appendix. The
triple-race that appears in the intermediate step can either be a BMN
or a BPS operator. Let us first consider the former case. We need to
compute, 
\bea
G^{23}_{my;pst}G^{23}_{pst,m'y'}&=&\left(y^{3/2}G^{12}_{m;ps/y}(\de_{y,s+t}+\de_{y,1-t})
+\frac{g_2}{\sqrt{J}}\de_{mp}\de_{ys}\sqrt{(1-y)t\tilde{t}}\right)\nonumber\\
{}&&\times\left({y'}^{3/2}G^{12}_{m';ps/y'}(\de_{y',s+t}+\de_{y',1-t})
+\frac{g_2}{\sqrt{J}}\de_{m'p}\de_{y's}\sqrt{(1-y')t\tilde{t}}\right),\nonumber
\eea
where one sums over $pst$. When last term in the first parenthesis
goes with the last term of the second we have the expression,
\be\la{eq}\frac{g_2^2}{J}\sum_{p=-\infty}^{\infty}\de_{pm}\de_{pm'}(J\int_0^1\ud
s)(\frac{J}{2}\int_0^{1-y}\ud
t)\de_{sy}\de_{sy'}\sqrt{(1-y)(1-y')}t(1-y-t)
=\de_{yy'}\de_{mm'}\frac{(1-y)^4}{24},\ee  
where in the integral over $t$ we divided by a factor of 2 to
reconcile with the double-counting (note that $t\to 1-t$ is not
distinguishable at the level of Feynman diagrams when triple trace is
BPS, and one should divide out the symmetry factor). 
When first term in the first parenthesis goes with second or third
terms of the second, both of the integrals over $s$
and $t$ are constraint by the delta-functions and one gets a $1/J$
suppression. A similar remark apply the case when second goes with
third. Therefore we see that all cross terms are suppressed and only
other non-vanishing contribution comes by matching second with second and third with
third. This is, 
$$\half 2(yy')^{3/2}\de_{yy'}\sum_{p=-\infty}^{\infty}\int_0^y\ud s
G^{12}_{m,ps/y}
G^{12}_{m',ps/y'}=y^4\sum_{p=-\infty}^{\infty}\int_0^1\ud x G^{12}_{m,px} G^{12}_{m',px}$$
where we again divided out a similar symmetry factor. It is easy to
see that when the intermediate triple-trace operators are BPS type one
gets the expression,
$$y^4\sum_{p=-\infty}^{\infty}\int_0^1\ud x G^{12}_{m,x}
G^{12}_{m',x}$$
instead of the above expression. Adding these two up and using
(\ref{id0}), one gets $2y^4G^{11}_{mm'}$. Combining it with the
contribution from (\ref{eq}) and comparing with (\ref{Cii2}) for the case
of $i=2$ we proved (\ref{id2}) at the leading
order. 

Next, we shall present two new ``interacting level'' decomposition
identities which are essentially the analogs of the identities given
in Appendix D of \cite{CFHM}:
\bea
\sum_{p,y'}\frac{p}{y'}G_{n,py'}G_{py',myz}&=&(n+\frac{m}{y})G_{n,myz},\la{intdec1}\\
\sum_{p,y'}\frac{p^2}{{y'}^2}G_{n,py'}G_{py',myz}&=&(n^2+(\frac{m}{y})^2)
G_{n,myz}+B^{13}_{n,myz}\la{intdec2}
\eea
where $B^{13}$ is given in (\ref{B13}). These identities can easily be
proven by the methods described above. 

As an application of these decomposition identities let us prove
(\ref{zeroine}). For notational simplicity we will no show the indices
fully in the following \eg we denote $\G_n^{myz}$ as $\G_1^3$,
etc. Eq. (\ref{dec}) gives,
\bea
\G_1^3&=&G^{33}\G_{31}+G^{32}\G_{21}+G^{31}\G_{11}\nonumber\\
{}&=&\half\frac{m}{y}(\frac{m}{y}-n)G^{13}+\half B^{13} 
-\half\sum_{p,y'}\frac{p}{y'}(\frac{p}{y'}-n)G^{12}G^{23}.\nonumber
\eea
where the second line follows after trivial algebra. Now, using
(\ref{id1}), (\ref{intdec1}) and (\ref{intdec2}) it is immediate to see that 
$$\G^3_1=0.$$

Let us now describe the evaluation of (\ref{sum11}) in section
2. We separate the LHS of (\ref{sum11}) into two parts as, 
\be\la{eq2}
\sum_p(-nA(p)+B(p))
\equiv\sum_p\left(-n\frac{\sin^2(\pi p
    y/y')}{(n-p/y')^2(n^2-(p/y')^2)}
+\frac{\sin^2(\pi p
    y/y')}{(n-p/y')^3}\right).
\ee
Evaluation of $\sum_p B(p)$ is easier. It can be written as, 
$$\sum_p B(p)=\frac{{y'}^3}{2}\left(-\sum_{p=-\infty}^{\infty}\frac{1}{(p-a)^3}
+\half\frac{d^2}{d
  a^2}\left\{2a\sum_{p=1}^{\infty}\frac{\cos(px)}{p^2-a^2}-\frac{1}{a}\right\}\right).$$
Each of the sums can be found in standard tables such as \cite{GR} and
the result is,
\be\la{B}
\sum_pB(p)=\pi^3y^2y'\cot(\pi ny').
\ee
To compute $\sum_p A(p)$ we write it as,
$$A(p)=\frac{{y'}^4}{2}\left(\frac{\cos(px)}{(p-a)^2(p^2-a^2)}-\frac{1}{(p-a)^2(p^2-a^2)}\right).$$
Second can be done by a standard computer code. First can be written
as, 
$$\sum_p\frac{\cos(px)}{(p-a)^2(p^2-a^2)}=-\frac{1}{a^4}
+(\frac{d^2}{dx^2}+a^2)\frac{d^2}{db^2}
\sum_{p=1}^{\infty}\frac{\cos(px)}{(p^2-b)}$$
where $b=\sqrt{a}$. This can be looked up in \cite{GR}. Combining the
result with (\ref{B}) one obtains (\ref{sum11}).  
 
\section{Computation of $G^{13}$, $G^{23}$, $\G^{13}$ and $\G^{23}$}

We will first describe the evaluation of free planar single-triple
correlator at the planar level,
$\<\bar{O}_n^J:O_m^{J_1}O^{J_2}O^{J_3}:\>$. We will refer to the 
operators that appear in this expression as 
``big operator'', ``operator 1'',  ``operator 2'' and 
``operator 3'', respectively. Let us also denote the ratios of the
``sizes'' of these operators by 
\be\la{ratio}
y=\frac{J_1}{J},\;\;z=\frac{J_2}{J},\;\;\tilde{z}=\frac{J_3}{J}.
\ee
Since the space-time dependence of two-point functions of scalar
operators is trivial we will only be interested in the coefficient
that multiplies the space-time factors, \ie $G^{13}$ and
$\G^{13}$. Nevertheless, let us show the space-time factors here, for
completeness. 
For the free case it is just product of $J+2$ scalar propagators,
$$\frac{1}{(4\pi^2x^2)^{J+2}}.$$
In case of one-loop interactions, one needs to perform the following
interaction over the position of the vertex, 
$$
\frac{1}{16\pi^4}
\int \frac{\ud^4 y}{y^4(y-x)^4}=\frac{\ln (\Lambda^2x^2)}{8\pi^2
  x^4}.$$
Therefore the space-time dependence at $\O(\lm')$ is, 
$$\frac{1}{8\pi^2(4\pi^2x^2)^{J+2}}\ln(\Lambda^2x^2).$$ 
Let us now describe the evaluation of the coefficients that multiply
these space-time factors. 

\begin{figure}[htb]
\centerline{ \epsfysize=6cm\epsffile{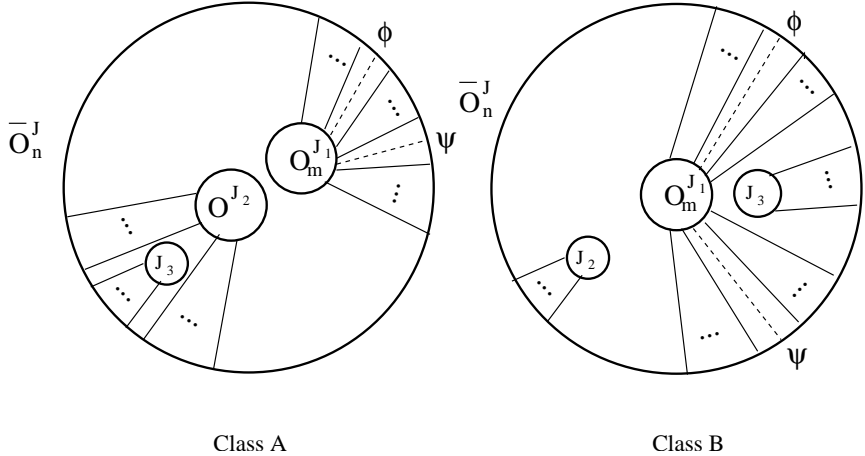}}
\caption{Two different classes of free diagrams. Dashed lines denote
  propagators of impurity fields.}
\end{figure}

General strategy is first to fix the
position of one operator, say 2 inside the big operator. Then we are left
with phase sums over positions of both of the impurities and the
position of operator 3 inside the big operator. Of course one still has to
take into account the cyclicity of 2 and 3 which yield a
multiplicative factor of $J_2J_3$. After fixing the position of 2, we
can divide the planar diagrams into two classes. 

In {\bf class A} (see Fig.10.)
operator 2 is 
``outside'' the bunch of lines connecting operator 1 to operator the
big operator, hence the phase sum over $\phi$ and $\psi$ is trivial:
$$J_1^2\int_0^1\ud a\int_0^1\ud b
e^{2i\pi(m-ny)a}e^{-2i\pi(m-ny)b}=\frac{\sin^2(2\pi
  ny)}{\pi^2(ny-m)^2}.$$
One also has to sum over the position of operator 3 ``under'' 2 and
position of operator 2 ``under'' 3. This gives a combinatorial factor
of $J_2+J_3$. As apparent from Fig.10. class A is equivalent to
single-double correlator up to the aforementioned overall
combinatorial factor. Therefore the left diagram in Fig.10. equals,
\be\la{free1}
G^{13}_A=\frac{g_2^2}{J}(1-y)\sqrt{\frac{z\tilde{z}}{y}}\frac{\sin^2(2\pi
  ny)}{\pi^2(n-k)^2}
\ee
where we took into account the normalization of operators and defined
$k\equiv m/y$. 

In {\bf class B} (see Fig.10), operator 3 is inserted inside the
bunch of lines connecting 1 to the big operator, hence the phase
summations become non-trivial. One fixes the position of 2 inside the
bunch, then sums over the positions of $\phi$ and $\psi$. As one
impurity jumps over operator 3 one gets an enhancement in the phase of
the big operator by a factor of $-2\pi in\tilde{z}$. One evaluates the sums
taking this point into account, than one sums over the position of
operator 2. This procedure gives, 
\bea\la{free2}
G^{13}_B&=&\frac{g_2^2}{\pi^2 J}\sqrt{\frac{z\tilde{z}}{y}}
\frac{1}{(n-k)^2}
\bigg(y(\sin^2(\pi n z)+\sin^2(\pi n
\tilde{z}))\nonumber\\
{}&&-\frac{1}{2\pi(n-k)}
(\sin(2\pi n y)+\sin(2\pi n z)+\sin(2\pi n \tilde{z}))\bigg).
\eea
Adding up (\ref{free1}) and (\ref{free2}) gives
eq. (\ref{C13})\footnote{We thank Neil Constable who
  computed this quantity by a completely different method (direct
  evaluation of the traces over the color structure and extracting out
  the $\O(g_2^2)$ piece) and who obtained the same result.} 

There are two consistency checks that one can perform. First of all -as
apparent from the diagrams- the final expression should be
symmetric in $J_2\leftrightarrow J_3 $.(\ref{C13}) nicely passes this
test. A more non-trivial test is to check whether $G^{13}$ reduces to
$G^{12}$ as one takes $J_3\to 0$. Straightforward algebra shows that, 
$$G^{13}_{n,myz}\to\sqrt{\frac{z}{J}}G^{12}_{n,my}$$
and confirms our expectation. 
    
Now let us discuss how to add interactions to Figs.10, by preserving
{\emph{planarity}}. As already mentioned for the evaluation of general
correlators in Appendix A, there are three distinct classes of planar
interactions: {\emph{contractible, semi-contractible and
    non-contractible.} Above we noted that evaluation of class
  A diagrams are completely equivalent to single-double
  correlator. This continues to be the case when one introduces
  planar interactions. There are only contractible and
  semi-contractible contributions in this case and since $\O(\lm')$ corrections to this
  correlator was already computed in \cite{CFHM} (that we reproduced in
  eq. (\ref{G12})), we will only show the final result,
\be\la{int1}
Class\;A\;interactions\;\Rightarrow\frac{g_2^2\lm'}{J\pi^2}(1-y)\sqrt{\frac{z\tilde{z}}{y}}\frac{\sin^2(2\pi
  ny)}{\pi^2(n-k)^2}(k^2-nk+n^2).
\ee

Let us explain the evaluation of interactions in class B in some
detail. Contractible interactions are coming from the situation where
an impurity interacts with its nearest-neighbor in such a way that
both interaction loops are contractible. As described in
Appendix A this gives a phase factor of 
\be\la{phasesupp}
(1-e^{\frac{-2i\pi n}{J}})(1-e^{\frac{2i\pi m}{J_1}})\approx
\frac{4\pi^2}{J^2}nk\ee
for each possible nearest-neighbour interaction. One should sum over
the insertions of this interaction between all adjacent line pairs
between operator 1 and the big operator in Fig.10.b, {\emph{except}} 
the particular position when this line pair coincides with the
position of operator 3. In this particular case one gets a
semi-contractible 
diagram (see Fig.11.a). 
This sum procedure obviously gives the phase factor in
(\ref{phasesupp}) times (\ref{free2}). 

\begin{figure}[htb]
\centerline{ \epsfysize=6cm\epsffile{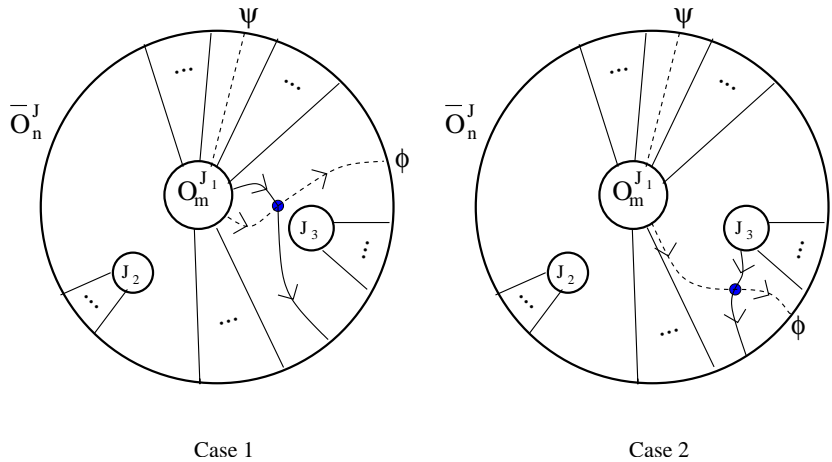}}
\caption{Two different types of semi-contractible diagrams. One should
  also consider the cases when two incoming and outgoing lines are
  interchanged. Also analogous contributions come from exchanging
  operator 2 with operator 3.}
\end{figure}

Semi-nearest interactions in class B arise in two possible ways. First
possibility is already mentioned above and shown in Fig.11.a. Another
possibility arise when one of the interaction loops is
non-contractible for another reason: the line pair that is incoming to the vertex
connect to different operators. This is illustrated in Fig.11.b. Evaluation
of the phase factors in both of these cases is simple: 

In case 1, when
one takes all possible orderings of the interacting pairs of lines, one
gets a factor of 
$$(1-e^{2i\pi n\tilde{z}})(1-e^{-\frac{2i\pi m}{J_1}})$$ in place of
(\ref{phasesupp}). Next, 
one has to sum over all possible positions of operator 3 in
Fig.11.a. Finally one evaluates the phase sum over $\psi$. There is an 
analogous contribution where $\psi$ impurity takes place in the
interaction instead of $\phi$. But that is obviously obtained from the former
just by taking the complex conjugate. 

In case 2, Fig.11, one has to sum over
the possibilities where $\phi$ interacts with the leftmost and the
rightmost $Z$ line in operator 3. Considering also the two different
orderings of the $\phi$ and $Z$ that are outgoing from the vertex, one
obtains an overall phase factor of, 
$$(1-e^{\frac{-2i\pi n}{J}})(1-e^{2i\pi n\tilde{z}}).$$
Next, just as in the case 1 above, one sums over all possible
insertions of operator 3 and positions of $\psi$ impurity. Similarly
one considers the conjugate case where $\psi$ takes place in the
interaction instead of $\phi$. Finally one gets similar expressions to
the ones obtained in case 1 and case 2 by exchanging the roles played
by operator 3 and operator 2.   

Combining all of the results above, namely both contractible and
semi-contractible contributions in class A given in (\ref{int1}),
contractible contributions in class B and all semi-nearest
contributions in class B, one obtains a surprisingly simple
expression. All of the factors conspire to give, 
\be\la{int2}
\G^{13}_A+\G^{13}_{B,cont.}+\G^{13}_{B,semi-cont.}=
\lm'(n^2-nk+k^2)G^{13}.
\ee
A few observations are in order. Notice that one obtains the same
form for the interacting single-double correlator as shown in
\cite{CFHM}, see eq. (\ref{G12}). The proportionality to
$G^{12}$ (or $G^{13}$ in our case) is obvious from the
beginning. Because eventually, the effect of interactions is 
to dress the free expression with an overall phase factor. The
surprise is that this phase factor, 
$$n^2-nk+k^2,$$
is the same in the cases of single-double and single-triple
correlators! One appreciates the non-triviality of this after seeing
the delicate conspiring of many different terms in our case. We also
see the same phase dependence, in case of double-triple correlator,
eq. (\ref{G23}). It is
tempting to believe that this remains to be true in case of general 
$i$-trace $j$-trace correlator. Namely we believe that the result of
contractible and semi-contractible interactions at $\O(\lm')$ for 
more general extremal correlators can be summarized as, 
$$\G^{i,j}_{cont.+semi-cont.}=\lm'(n^2-nk+k^2)G^{i,j}.$$
Actually it suffices to see this behaviour in case of extremal
correlators of the type $\G^{1,i}$ since $\G^{i,j}$ can be related to
this by the disconnectedness argument.  

However, this is not the whole story. There is a very important new
class of planar diagrams which
contributes to $G^{13}$: {\emph{non-contractible}} diagrams. This was
absent in the case of $\G^{12}$ because there was not enough number of
operators to create this new interaction topology. This will become
clear in the following. 

\begin{figure}[htb]
\centerline{\epsfxsize=12cm \epsfysize=12cm\epsffile{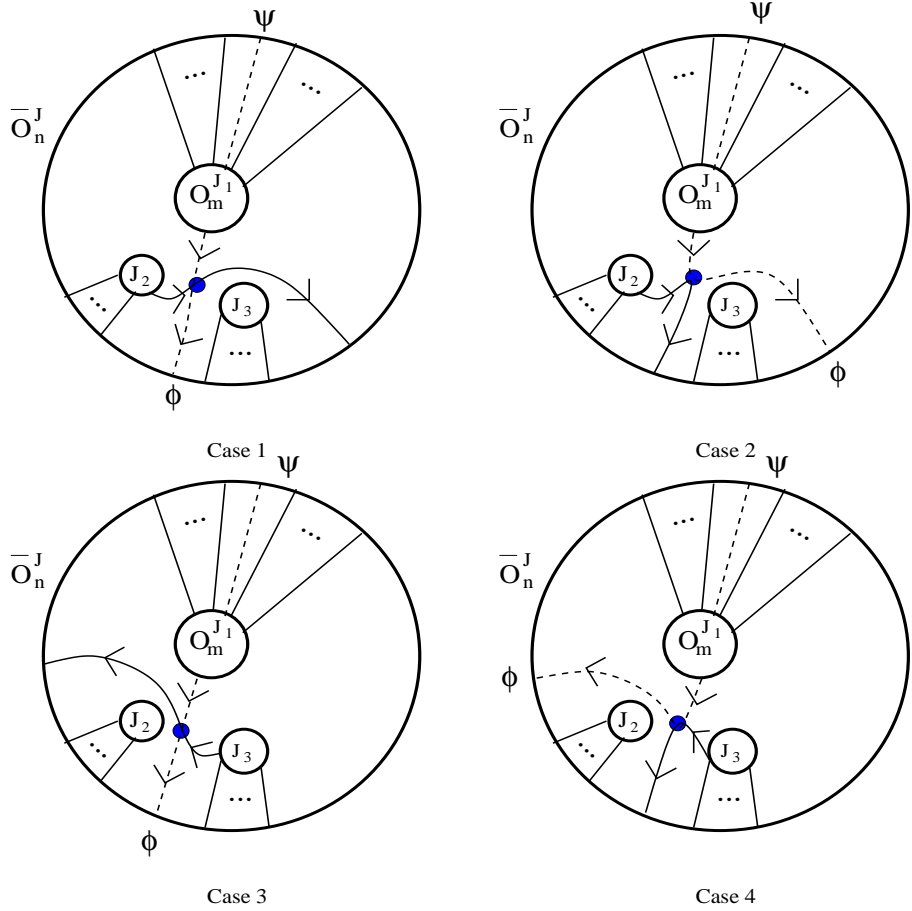}}
\caption{Contributions form non-contractible planar interactions. There
are four more diagrams which are obtained by exchanging operator 2
with operator 3.}
\end{figure}

We show all possible non-contractible diagrams in Fig.12. Note that
both of the interaction loops are non-contractible in this case. The
loop formed by incoming lines is non-contractible because they belong
to different operators. The loop formed by the outgoing lines is
non-contractible because there is an operator inserted between
them. This exemplifies our schematic discussion about the
non-contractibility of planar diagrams in Appendix A where we referred
to these possibilities as case 1 and 2. At first sight one expects
that these diagrams be suppressed by a factor of $1/J^2$ when compared
with the contractible diagrams or by a factor of $1/J$ when compared
with the semi-contractible diagrams, because the sums over the position of
the operator 3 and the position of $\phi$ impurity are missing. However
as noted in the general discussion of Appendix A, there is a
compensating enhancement coming from the overall phase factors
associated with these diagrams, namely the $O(1/J^2)$ phase
suppression given by (\ref{phasesupp}) is absent. Therefore these diagrams
are on the equal footing with the rest \ie (\ref{int2}). 

The evaluation of non-contractible diagrams is the simplest. One adds
up all possible contributions that are displayed in Fig.12, and
include the analogous cases where one interchanges operator 2
with operator 3. Finally one performs the phase summation over
$\psi$. Adding this result with the conjugate one which is obtained
by interchanging the roles of $\phi$ and $\psi$, one gets
(\ref{B13}). 

Adding up (\ref{B13}) with (\ref{int2}), one obtains the total
$\O(g_2^2\lm')$ single-triple correlator, (\ref{G13}). 
Again, there are two consistency checks that one can perform. Firstly
it is easy to see that, (\ref{G13}) is symmetric in
$J_2\leftrightarrow J_3$. Secondly when one takes the limit $J_3\to
0$, $B^{13}$ vanishes (as it should) and the rest of the expression
boils down to the single-double result, 
$$\G^{13}\to \sqrt{\frac{z}{J}} \G^{12}.$$

Let us now explain the computations that lead to the expressions in
(\ref{C23}) and (\ref{G23}). 
When compared with the evaluation of single-triple correlators,
evaluation of lowest order $G^{23}$ and $\G^{23}$ is almost trivial. This is
because only the disconnected diagrams contribute to these
correlators at $\og2$. We now describe the evaluation of $G^{23}$. 
It will suffice to describe possible ways that
$2\to 3$ correlator can be separated into $1\to 1$ and $1\to
2$. Consider the correlator  
$\<:\bar{O}^{J_1}_n\bar{O}^{J_2}::O^{J_3}_m
O^{J_4}O^{J_5}:\>$ define the ratios of lengths of the operators, 
\be
y=\frac{J_1}{J},\;\;y'=\frac{J_3}{J},\;\;z'=\frac{J_4}{J},\;\;\tilde{z'}=\frac{J_5}{J}.
\ee

Since the impurity fields in operator 1 and
operator 3 should be Wick contracted with each other, the only disconnected
contributions arise when, 
\begin{enumerate}
\item 1 connects to 3, 2 connects to 4 and 5, 
\item 1 connects to 3 and 4, 2 connects to 5, 
\item 1 connects to 3 and 5, 2 connects to 4.
\end{enumerate}
A simple loop counting shows that all other contractions will result
in higher orders in $g_2$. Cases 2 and 3 are easily expressible in
terms of the results already reported in the literature, (see
\cite{Constable}). Therefore we only show case 1 which turns out to be
the simplest, 
$$\<:\bar{O}^{J_1}_n\bar{O}^{J_2}::O^{J_3}_m
O^{J_4}O^{J_5}:\>_1=\<\bar{O}^{J_1}_n O^{J_3}_m\>\<\bar{O}^{J_2}:O^{J_4}O^{J_5}:\>,$$
where one needs the lowest order contributions to the correlators on
RHS. First one is just $\de_{nm}$. One evaluates the BPS correlator
above by noting that the cyclicity factor of $J_2J_4J_5$ and the
normalizations; hence one gets, 
$$\frac{g_2}{\sqrt{J}}\sqrt{(1-y)z'\tilde{z}'}.$$
Combining this contribution with cases 2 and 3, one easily obtains,
(\ref{C23}). 

Computation of $\G^{23}$ at the lowest order, $\O(g_2\lm')$, goes by
inserting planar interactions into the cases 1,2, and 3 that we listed
above in all possible
ways. In case 1, interactions can only be inserted in the correlator
1-3 since 2-4+5 -being a BPS corrrelator- does not receive radiative
corrections. For the same reason interactions can be inserted only in
the first correlators in cases 2 and 3. Necessary computations were
already done in the literature (see \eg \cite{Constable}, \cite{CFHM})
and one immediately gets, (\ref{G23}).

\end{document}